\documentclass[prb,twocolumn,showpacs,preprintnumbers,amsmath,aps]{revtex4}
\usepackage{graphicx,hyperref}
\usepackage{bm}
\usepackage{subfigure}

\newcommand\des{{\hbox{\begin{scriptsize}d\ed{scriptsize}}}}

\newcommand\etab{\bar \eta}

\newcommand\cusp{\text{cusp}}
\newcommand\DR{\text{DR}}
\newcommand{\dr}{{\text{DR}}}
\renewcommand{\cusp}{{\text{cusp}}}

\def\nbC{{\mathchoice {\setbox0=\hbox{$\displaystyle\rm C$}%
\hbox{\hbox to0pt{\kern0.4\wd0\vrule height0.9\ht0\hss}\box0}}
{\setbox0=\hbox{$\textstyle\rm C$}\hbox{\hbox
to0pt{\kern0.4\wd0\vrule height0.9\ht0\hss}\box0}}
{\setbox0=\hbox{$\scriptstyle\rm C$}\hbox{\hbox
to0pt{\kern0.4\wd0\vrule height0.9\ht0\hss}\box0}}
{\setbox0=\hbox{$\scriptscriptstyle\rm C$}\hbox{\hbox
to0pt{\kern0.4\wd0\vrule height0.9\ht0\hss}\box0}}}}
\def\nbQ{{\mathchoice {\setbox0=\hbox{$\displaystyle\rm
Q$}\hbox{\raise
0.15\ht0\hbox to0pt{\kern0.4\wd0\vrule height0.8\ht0\hss}\box0}}
{\setbox0=\hbox{$\textstyle\rm Q$}\hbox{\raise
0.15\ht0\hbox to0pt{\kern0.4\wd0\vrule height0.8\ht0\hss}\box0}}
{\setbox0=\hbox{$\scriptstyle\rm Q$}\hbox{\raise
0.15\ht0\hbox to0pt{\kern0.4\wd0\vrule height0.7\ht0\hss}\box0}}
{\setbox0=\hbox{$\scriptscriptstyle\rm Q$}\hbox{\raise
0.15\ht0\hbox to0pt{\kern0.4\wd0\vrule height0.7\ht0\hss}\box0}}}}
\def\nbT{{\mathchoice {\setbox0=\hbox{$\displaystyle\rm
T$}\hbox{\hbox to0pt{\kern0.3\wd0\vrule height0.9\ht0\hss}\box0}}
{\setbox0=\hbox{$\textstyle\rm T$}\hbox{\hbox
to0pt{\kern0.3\wd0\vrule height0.9\ht0\hss}\box0}}
{\setbox0=\hbox{$\scriptstyle\rm T$}\hbox{\hbox
to0pt{\kern0.3\wd0\vrule height0.9\ht0\hss}\box0}}
{\setbox0=\hbox{$\scriptscriptstyle\rm T$}\hbox{\hbox
to0pt{\kern0.3\wd0\vrule height0.9\ht0\hss}\box0}}}}
\def\nbS{{\mathchoice
{\setbox0=\hbox{$\displaystyle     \rm S$}\hbox{\raise0.5\ht0%
\hbox to0pt{\kern0.35\wd0\vrule height0.45\ht0\hss}\hbox
to0pt{\kern0.55\wd0\vrule height0.5\ht0\hss}\box0}}
{\setbox0=\hbox{$\textstyle        \rm S$}\hbox{\raise0.5\ht0%
\hbox to0pt{\kern0.35\wd0\vrule height0.45\ht0\hss}\hbox
to0pt{\kern0.55\wd0\vrule height0.5\ht0\hss}\box0}}
{\setbox0=\hbox{$\scriptstyle      \rm S$}\hbox{\raise0.5\ht0%
\hboxto0pt{\kern0.35\wd0\vrule height0.45\ht0\hss}\raise0.05\ht0%
\hbox to0pt{\kern0.5\wd0\vrule height0.45\ht0\hss}\box0}}
{\setbox0=\hbox{$\scriptscriptstyle\rm S$}\hbox{\raise0.5\ht0%
\hboxto0pt{\kern0.4\wd0\vrule height0.45\ht0\hss}\raise0.05\ht0%
\hbox to0pt{\kern0.55\wd0\vrule height0.45\ht0\hss}\box0}}}}
\def\nbZ{{\mathchoice {\hbox{$\sf\textstyle Z\kern-0.4em Z$}}
{\hbox{$\sf\textstyle Z\kern-0.4em Z$}}
{\hbox{$\sf\scriptstyle Z\kern-0.3em Z$}}
{\hbox{$\sf\scriptscriptstyle Z\kern-0.2em Z$}}}}

\begin{document}

\title{Fixed points and their stability in the functional renormalization group of random field models}

\author{Maxime Baczyk} \email{baczyk@lptmc.jussieu.fr}
\affiliation{LPTMC, CNRS-UMR 7600, Universit\'e Pierre et Marie Curie,
bo\^ite 121, 4 Pl. Jussieu, 75252 Paris c\'edex 05, France}

\author{Gilles Tarjus} \email{tarjus@lptmc.jussieu.fr}
\affiliation{LPTMC, CNRS-UMR 7600, Universit\'e Pierre et Marie Curie,
bo\^ite 121, 4 Pl. Jussieu, 75252 Paris c\'edex 05, France}

\author{Matthieu Tissier} \email{tissier@lptmc.jussieu.fr}
\affiliation{LPTMC, CNRS-UMR 7600, Universit\'e Pierre et Marie Curie,
bo\^ite 121, 4 Pl. Jussieu, 75252 Paris c\'edex 05, France}

\author{Ivan Balog} \email{balog@ifs.hr}
\affiliation{LPTMC, CNRS-UMR 7600, Universit\'e Pierre et Marie Curie,
bo\^ite 121, 4 Pl. Jussieu, 75252 Paris c\'edex 05, France}
\affiliation{Institute of Physics, P.O.Box 304, Bijeni\v{c}ka cesta 46, HR-10001 Zagreb, Croatia}

\date{\today}

\begin{abstract}
  We consider the zero-temperature fixed points controlling the
  critical behavior of the $d$-dimensional random-field Ising, and
  more generally $O(N)$, models.  We clarify the nature of these fixed
  points and their stability in the region of the $(N,d)$ plane where
  one passes from a critical behavior satisfying the $d\rightarrow
  d-2$ dimensional reduction to one where it breaks down due to the
  appearance of strong enough nonanalyticities in the functional
  dependence of the cumulants of the renormalized disorder. We unveil
  an intricate and unusual behavior.

\end{abstract}


\maketitle

\section{Introduction}

In a recent series of papers,\cite{tarjus04,tissier06,tissier06b,tissier11,baczyk13} we have shown how the critical behavior of the 
$d$-dimensional random-field Ising, and more generally $O(N)$, models can be fully described through the functional renormalization 
group (FRG). We have in particular stressed that the solution to many puzzles associated with this critical behavior lies in the 
existence of a transition between a region satisfying the $d\rightarrow d-2$ dimensional reduction, \textit{i.e.} where the critical 
behavior of the random-field system is identical to that of the corresponding pure model in two dimensions less,\cite{aharony76,grinstein76,young77,parisi79} 
and one where dimensional reduction is broken.

The above transition takes place at a nontrivial location in the $(N,d)$ plane, in contrast for instance with the case of an interface in a 
random environment where dimensional reduction is always wrong below the upper critical dimension.\cite{fisher86b,narayan92,
FRGledoussal-chauve,FRGledoussal-wiese} For the random-field $O(N)$ model [RF$O(N)$M], the zero-temperature fixed point 
associated with dimensional reduction disappears below a line $d_{\DR}(N)$ that is close to $5.1$ when $N=1$ and decreases as 
$N$ increases, reaching $d=4$, the lower critical dimension for ferromagnetism in the presence of a continuous $O(N)$ symmetry, 
when $N=18$.\cite{tissier06,tissier11} Below this line, the zero-temperature fixed point controlling the critical behavior is characterized 
 by strong enough nonanalyticities in the functional dependence of the cumulants of the renormalized disorder. The latter take the form 
 of a linear ``cusp'' in the cumulants of the renormalized random field. Physically, this results from the presence of large-scale 
 collective events known as ``avalanches'' (whose fractal dimension is then equal to the fractal dimension of the total 
 magnetization at criticality)\cite{tarjus13}. Formally, this leads to a failure of the Ward-Takahashi identities associated with the 
 underlying supersymmetry of the model\cite{parisi79} and to a breaking of the latter.\cite{tissier11}

 In this paper, we take a closer look at the transition from the
 regime controlled by the dimensional-reduction fixed point to that
 controlled by the ``cuspy'' fixed point. Although this takes place in
 an unphysical region in systems with short-range interactions and
 disorder correlations (but could nonetheless be studied in $d=3$ when
 one allows for long-range interactions and disorder correlations as
 we have recently pointed out\cite{baczyk13}), the issue is important
 to underpin the whole FRG-based description.

A first question is whether one can find, within the FRG, operators that become relevant as dimensional reduction breaks down. To this end 
we have studied in more detail the stability of the ``cuspless'' fixed point associated with dimensional reduction  to a ``cuspy'' perturbation,
\textit{i.e.} a perturbation displaying a linear cusp in the cumulants of the renormalized random field, when $d\geq d_{DR}$. 
Quite surprisingly, we find two different mechanisms for the appearance and disappearance of the stable (critical) fixed point, depending 
on the value of $N$ (or $d$).  

For $N$ sufficiently large, the cuspless fixed point becomes unstable
with respect to a cuspy perturbation, and this occurs at a nontrivial
dimension $d_{\cusp}(N)$ that is close to, but different from,
$d_{\DR}(N)$. As a result, there is a range of dimensions
$d_{\DR}(N)<d<d_{\cusp}(N)$ [or, alternatively, of number of
components, $N_{\DR}(d)<N<N_{\cusp}(d)$] for which models described by
a cuspless initial condition flow at criticality to the cuspless fixed
point associated with dimensional reduction whereas models already
described by a cuspy initial condition flow to a cuspy fixed point for
which dimensional reduction fails. In an enlarged space of functions
including those with a linear cusp, only the latter fixed point is
fully stable (except for the usual relevant direction needed to tune
the critical-point condition).

For a threshold value $N_{\rm x}$, and correspondingly a threshold dimension $d_{\rm x}=d_{DR}(N_{\rm x})=d_{\cusp}(N_{\rm x})$, 
the two critical lines $d_{DR}(N)$ and $d_{\cusp}(N)$ meet: see Fig. 1. For $N<N_{\rm x}$ and $d>d_{\rm x}$, the 
cuspless fixed point that governs the critical physics remains stable under cuspy perturbations down to $d=d_\DR(N)$, at 
which point it disappears. A cuspy fixed point then emerges continuously from the cuspless one, through a boundary-layer 
mechanism.

We derive the above results by a combination of approaches. We investigate the RF$O(N)$M near its
lower critical dimension, in $d=4+\epsilon$, through the perturbative
FRG in section \ref{sec:RFO(N)M}. In section \ref{sec:toy_model}, we
illustrate the mechanisms for the appearance of cuspy fixed points and
the disappearance of cuspless ones in a toy model inspired from the
beta function of the RF$O(N)$M. We finally address the short- and
long-range versions of the random-field Ising model (RFIM) through the
nonperturbative FRG in section \ref{sec:RFIM}.
\begin{figure}[tb]
  \centering
\includegraphics[width=.8 \linewidth]{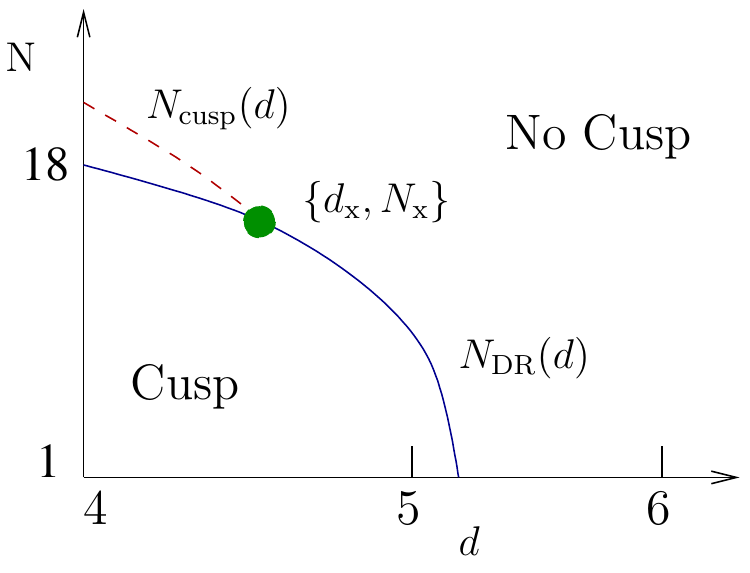}
\caption{Schematic phase diagram of the RF$O(N)$M in the $(N,d)$
  plane. The full line is $N_{DR}(d)$, where the cuspless fixed points
  present for $N>N_{DR}(d)$ disappear, and the dashed line is
  $N_{\cusp}(d)$, where the cuspless critical fixed point becomes
  unstable to a cuspy perturbation. The two lines meet at $N_{\rm
    x}\simeq 14$ and $d_{\rm x}\simeq 4.4$ (estimated from a 2-loop
  perturbative FRG in $d=4+\epsilon$). To the right of the threshold
  point, the cuspless critical fixed point disappears at $N_{DR}(d)$
  when it is still stable with respect to a cuspy perturbation. A
  stable cuspy fixed point then appears continuously but through a
  boundary-layer mechanism.}
\label{fig_phase_diagram}
\end{figure}

\section{The RFO(N)M in $d=4+\epsilon$}
\label{sec:RFO(N)M}

The long-distance physics of the RF$O(N)$M is described by the following Hamiltonian or bare action,
\begin{equation}
\begin{aligned}
\label{eq_ham_rfon}
&S[\bm{\varphi},\mathbf h]= \\& \int_{x} \bigg\{ \frac{1}{2}  \vert\partial \bm\varphi(x) \vert ^2 + \frac{\tau}{2}  
\vert \bm\varphi(x) \vert ^2  +  \frac{u}{4!} ( \vert \bm\varphi(x) \vert ^2)^2 -
\mathbf h(x)\bm{.} \bm{\varphi}(x) \bigg\} ,
\end{aligned}
\end{equation}
where $ \int_{x} \equiv \int d^d x$,  $\bm\varphi(x)$ is an $N$ component field,  
and $\mathbf h(x)$ is a random source (a random magnetic field in the language of magnetic systems) with zero mean and a variance
$\overline{h^{\mu}(x)h^{\nu}(y)}=\delta_{\mu\nu}\, \Delta_B(x-y)$, where $\mu,\nu=1, \cdots,N$ and an overline denotes an average over 
the random field. For the usual short-range model, the function $\Delta_B(x-y)$ can be taken as $\Delta_B\, \delta(x-y)$. In section 
\ref{sec:RFIM}, we will  consider a version with both long-ranged disorder correlations $\Delta_B(x-y)\sim \vert x-y\vert^{-(d-\rho)}$ and 
long-ranged interactions leading to a kinetic term with a fractional laplacian in place of the standard $\vert\partial \bm\varphi(x) \vert ^2$ term 
above.\cite{baczyk13}

Near the lower critical dimension for ferromagnetism ($d=4$), the
critical behavior of the RF$O(N)$M is captured by a nonlinear-sigma
model that in turn can be studied through a perturbative but
functional RG. The resulting FRG flow equations have been obtained to
one\cite{fisher85,feldman01,tissier06,tissier06b} and two
loops.\cite{tissier06b,ledoussal06} The central quantity is the
renormalized second cumulant of the random field $\Delta_k(z)$ [noted
$R_k'(z)$ in previous work], where $k$ is the running infrared cutoff
and $z$ is the cosine of the angle between fields in two different
replicas of the system.\cite{tissier06,tissier06b} A linear cusp in
this parametrization corresponds to a term in $\sqrt{1-z}$ as
$z\rightarrow 1$. We then use the terminolgy ``cuspy'' to describe a
function with this behavior and ``cuspless'' if the function and its 
first derivative, $\Delta(1)$ and $\Delta'(1)$, in $z=1$  are finite.

For completeness we recall the FRG equation for $\Delta_k(z)$ at
one-loop, in $d=4+\epsilon$:
\begin{equation}
  \label{eq_flow_delta}
  \begin{split}
\frac 1\epsilon \partial_{t}& \Delta_{k}(z) =   \Delta_{k}(z) -  \bigg\lbrace (N-3) \Delta_{k}(1) \Delta_{k}(z) +z\Delta_{k}(z)^2 \\& +(N-3 + 4 z^2)
\Delta_{k}(z) \Delta_{k}'(z)- (N+1) z \Delta_{k}(1) \Delta_{k}'(z)  \\& - z (1- z^2) \Delta_{k}(z) \Delta_{k}''(z)  + ( 1 - z^2) \Delta_{k}(1) \Delta_{k}''(z)
-\\ & 3 z ( 1-z^2)   \Delta_{k}'(z)^2+ (1-z^2)^2 \Delta_{k}'(z) \Delta_{k}''(z) \bigg\rbrace
\end{split}
\end{equation}
where we have rescaled $\Delta_k$ by $8\pi^2\epsilon$. The RG 
``time'' $t$ is defined such that the long-distance physics is recovered
when $t\to +\infty$, \textit{i.e.} $t=\log(\Lambda/k)$ with $\Lambda$ the microscopic or ultraviolet scale. The two anomalous 
dimensions $\eta$ and $\bar\eta$ characterizing the spatial dependence of the correlation functions at criticality in random-field 
systems are expressed in terms
of $\Delta(1)$ at the fixed point:
\begin{align}
  \eta&=\epsilon \Delta_\star(1)\\
\etab&=\epsilon \big [-1+(N-1)\Delta_\star(1)\big ] \,.
\end{align}
For sufficiently large $N$, the critical behavior is controlled by a
fixed point at which $\Delta_*(z)$ has only a ``subcusp'', with a
leading nonanalytic behavior in $(1-z)^{\alpha_*(N)}$ and
$\alpha_*(N)>1$, which implies that $\Delta_k(1)$ and $\Delta_k'(1)$
remain finite during the flow. A direct calculation shows that, under
this hypothesis, the evolution of $\Delta_k(1)$ only depends on
$\Delta_k(1)$ and that of $\Delta_k'(1)$ only depends on $\Delta_k(1)$
and $\Delta'_k(1)$. At the corresponding cuspless fixed point, one
has\cite{fisher85,tissier06b}
\begin{equation}
  \label{eq_cuspless}
  \begin{split}
\frac{\Delta_*(1)}\epsilon&=\frac1{N-2}\\
\frac{\Delta'_*(1)}\epsilon&=\frac{ (N-8)-\sqrt{(N-2)(N-8)}}{2(N-2)(N+7)} \,.
  \end{split}
\end{equation}
The square root in the expression of $\Delta'_*(1)$ implies that this fixed point exists only for $N>N_\DR=18$.

The determination of $\alpha_*(N)$ is obtained as follows. Suppose that
the function $\Delta_k(z)$ has a leading singularity in $\alpha$ with
$\alpha>1$:
$\Delta(z)=\Delta_k(1)-\Delta'_k(1)(1-z)+\cdots+a_k(1-z)^\alpha+\cdots$. One 
can easily show that the flow of $a_k$ is linear in $a_k$ and that it
depends on $\Delta_k(1)$ and $\Delta_k'(1)$ only:
\begin{equation}
  \label{eq_flow_ak}
  \frac 1\epsilon\partial_t a_k=a_k\, \Lambda_{\alpha+1}(\Delta_k(1),\Delta_k'(1))
\end{equation}
with\cite{tissier06b}
\begin{equation}
  \label{eq_Lambda}
  \begin{split}
    \Lambda_{p} (\Delta(1),\Delta'(1)) =&  1 - \Delta(1) \big [ N(2 - p) + 2 p^2 + p - 4\big ] \\&- \Delta'(1) p ( 6 p + N - 5) \,.
  \end{split}
\end{equation}
The only way to have a nonvanishing amplitude for a subcusp with exponent $\alpha_*$ is that 
\begin{equation}
  \label{eq_def_alpha*}
  \Lambda_{\alpha_*
    +1}(\Delta_\star(1),\Delta_\star'(1))=0.  
\end{equation}
By using the fixed-point solution given in Eq. (\ref{eq_cuspless}) we then obtain an
explicit expression for $\alpha_*(N)$, which we do not reproduce here. It is
found that $\alpha_*(N)$ decreases as $N$ decreases until it reaches
$\alpha_*(N=18)=3/2$.\cite{tissier06,tissier06b} Below $N=18=N_\DR$, the only nontrivial fixed points have a linear cusp, with
$\alpha_*(N)=1/2$.

The eigenvalues describing the stability of the cuspless fixed point
under consideration are obtained by linearizing the FRG flow equations
around this fixed point. In our previous work we found that for $N\geq
18$ the cuspless fixed point described by Eq.~(\ref{eq_cuspless}) is
stable with respect to cuspless perturbations, except of course for
the relevant direction [here, $\Delta(1)$] that must be fine-tuned to
reach the critical point. Starting from a cuspless initial condition
for $\Delta_{k=\Lambda}(z)$ at the ultraviolet scale, one ends up
after fine-tuning the relevant parameter $\Delta(1)$ at a cuspless
fixed point, and the critical exponents are given by the
dimensional-reduction predictions. However, at the time, we did not
systematically investigate the stability of the cuspless fixed point
with respect to a \textit{cuspy} perturbation. This had been done by
Sakamoto et al.\cite{sakamoto06} in the large $N$ limit in an
expansion in $1/N$, at one- and two-loop orders. The outcome was that
the cuspless fixed point is then stable to all perturbations, with and
without a cuspy functional behavior.

\begin{figure}[tb]
  \centering
\includegraphics[width=\linewidth]{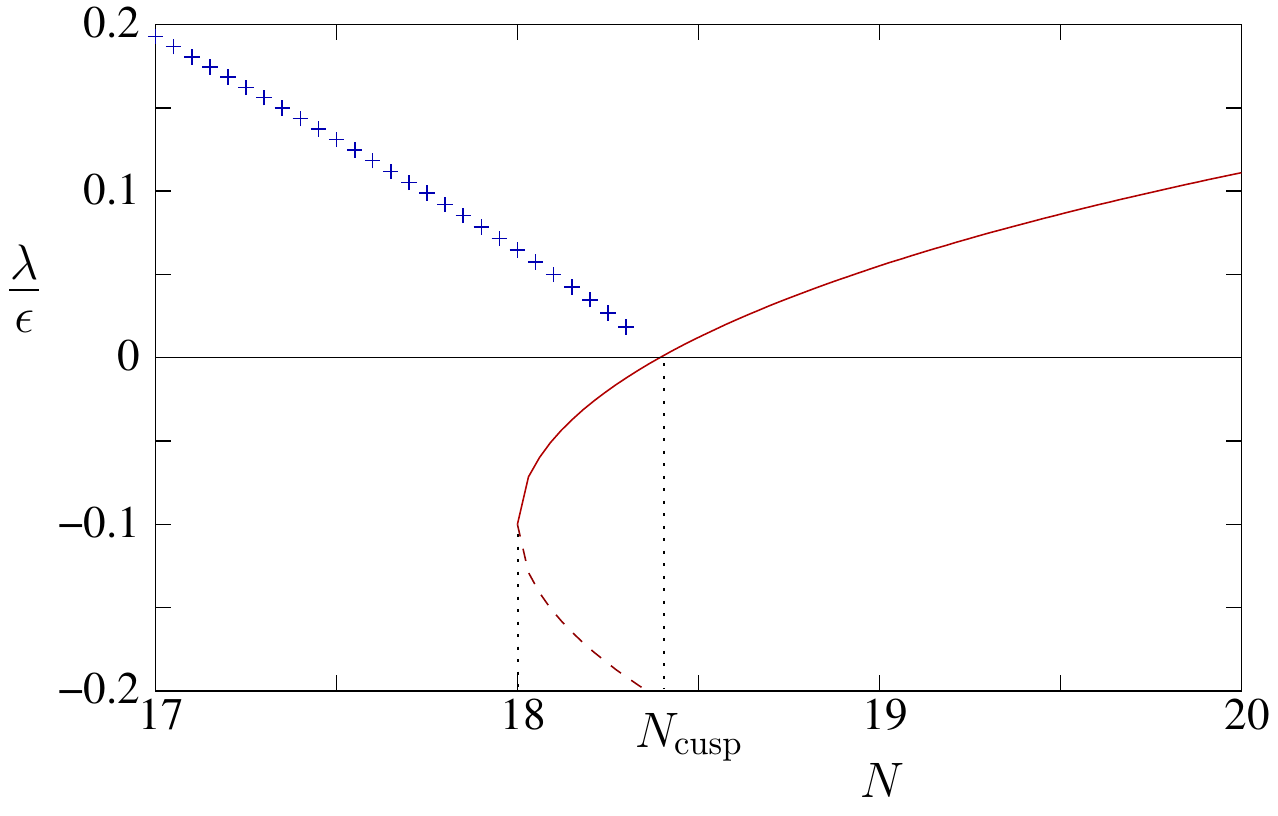}
\caption{RF$O(N)$M at one-loop order in $d=4+\epsilon$: Eigenvalue $\lambda(N)/\epsilon$ associated with a cuspy 
perturbation around  different fixed points. The full curve corresponds to the cuspless critical fixed point associated with 
dimensional reduction [Eq.~(\ref{eq_eigenvalueO(N)})], the dashed line corresponds to the unstable conjugate cuspless 
fixed point [Eq.~(\ref{eq_eigenvalueO(N)_ins})], and the crosses correspond to the cuspy fixed point.}
\label{fig_lambda}
\end{figure}

We are primarily interested in the eigenvalue $\lambda(N)$ associated
with a cuspy eigenfunction $f_N(z)$. It coincides with $\Lambda_{3/2}$ obtained from Eqs. (\ref{eq_Lambda}, \ref{eq_def_alpha*}) 
and reads:
\begin{equation}
\label{eq_eigenvalueO(N)}
\frac{\lambda(N)}{\epsilon}=\frac 1{4(N+7)}\left(3(N+4)\sqrt{\frac{N-18}{N-2}}-N+8\right)
\end{equation}
where a positive value means an irrelevant direction. One checks that the result of Ref. [\onlinecite{sakamoto06}] is recovered 
in the large $N$ limit: $\lambda(N)/\epsilon=1/2-9/(2N)-57/(2 N^2)+O(1/N^3)$.  We have plotted
the eigenvalue $\lambda(N)$ in Fig.~\ref{fig_lambda}. It decreases as
$N$ decreases, reaches zero when $N=N_{\cusp}=2(4+3\sqrt 3)\simeq
18.3923\cdots$ and then changes sign. (Its value in $N=18$ is equal to
$-1/10$.) The cuspless fixed point therefore becomes unstable with
respect to a cuspy perturbation at a value of $N$ which is slightly
larger than the value $N_{\DR}=18$ below which cuspless fixed points
no longer exist.

We have repeated the analysis for the other cuspless fixed point that
is somehow conjugate to the critical one described above but has one
more (cuspless) relevant direction.\cite{fisher85,tissier06b} It is characterized by
$\Delta_*(1)=\epsilon/(N-2)$ and $\Delta'_*(1)=\epsilon [(N-8)+
\sqrt{(N-2)(N-8)}]/[2(N-2)(N+7)]$. The two cuspless fixed point merge and disappear when $N=N_{\DR}=18$. The 
eigenvalue associated with a cuspy perturbation around this unstable cuspless fixed point is now given by
\begin{equation}
\begin{aligned}
\label{eq_eigenvalueO(N)_ins}
\frac{\lambda(N)}{\epsilon}=\frac 1{4(N+7)}\left[-3(N+4)\sqrt{\frac{N-18}{N-2}}-N+8\right ] \,.
\end{aligned}
\end{equation}
A cuspy perturbation is therefore a relevant direction from $N\rightarrow \infty$,
where it behaves as $-1+12/N-24/N^2+O(1/N^3)$, down to $N_{\DR}$. This
is also displayed in Fig.~\ref{fig_lambda}.

\begin{figure}[tb]
  \centering
 \includegraphics[width=.32\linewidth]{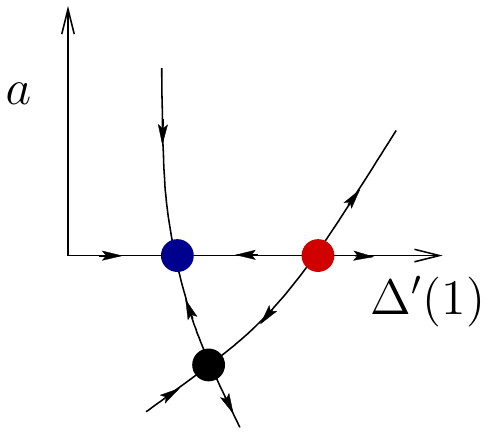}
 \includegraphics[width=.32\linewidth]{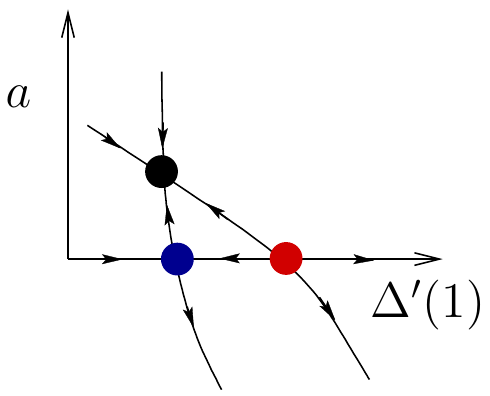}
 \includegraphics[width=.32\linewidth]{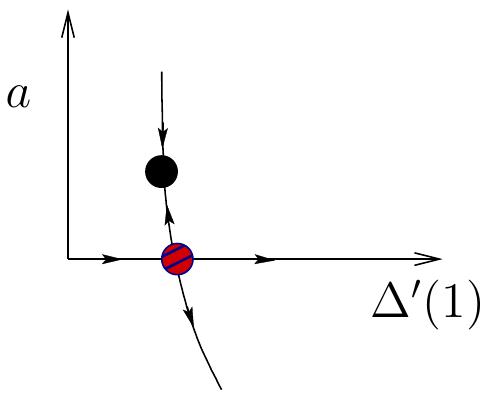}
 \caption{Schematic flow for the RFO($N$) model in the plane
   $(\Delta'(1),a)$ where $a$ represents the amplitude of the cusp.
   The blue and red points represent the stable and unstable analytic
   fixed points and the black point represents the cuspy fixed point.
   Left: for $N>N_\cusp$; middle: $N_\DR<N<N_\cusp$; right:
   $N=N_{DR}$. In the presence of a cusp ($a\neq 0$), $\Delta'(1)$
   should be interpreted as (minus) the coefficient of the linear term
   in $(1-z)$ when $z\to 1$.}
  \label{fig:fp}
\end{figure}

The destabilization of the cuspless critical (\textit{i.e.} stable)  fixed point at $N_\cusp$
occurs in a standard way. We find by a numerical integration
of the beta function that there exists a third fixed point, characterized by a cuspy functional form, which coincides 
with the cuspless critical fixed point for $N=N_\cusp$ and  is stable for $N<N_{\cusp}$. The second smallest
eigenvalue for this cuspy fixed point is also shown in Fig.~\ref{fig_lambda}. The general scenario for the exchange of stability 
of the fixed points is therefore quite common and appears in many other situations, such as for instance the
destabilization of the $O(N)$ Wilson-Fisher fixed point upon adding
anisotropic interactions.\cite{grinstein76b} We give in Fig.~\ref{fig:fp} a schematic description of the RG 
flows to illustrate the evolution of the different fixed points.

The previous discussion makes it clear that there is a small domain $N_{\DR}<N<N_\cusp$ where the critical behavior is described 
by the dimensional-reduction property if the initial condition of the flow is cuspless, 
but where it is governed by a cuspy fixed point and a breakdown of  dimensional reduction otherwise. One should keep in mind that the 
FRG framework considered here starts with a coarse-grained Landau-Ginzburg description of the system at the microscopic (ultraviolet) 
scale: see Eq.~(\ref{eq_ham_rfon}). As long as there exists only one fixed point and that we
limit our investigation to the critical physics, the detailed properties of the microscopic system are irrelevant. However, in the small
region between $N_{\DR}$ and $N_\cusp$, the situation is more intricate. A discussion of the $d=0$ (1-site) problem\cite{tissier11} shows 
that the presence of a cusp at the microscopic level, which is associated with avalanches, is most probably the rule rather than
the exception at $T=0$. This implies that physical systems at $T=0$ are likely to always flow to the cuspy fixed point when 
$N<N_\cusp$.\cite{footnote}

\begin{figure}[tb]
  \centering
\includegraphics[width=\linewidth]{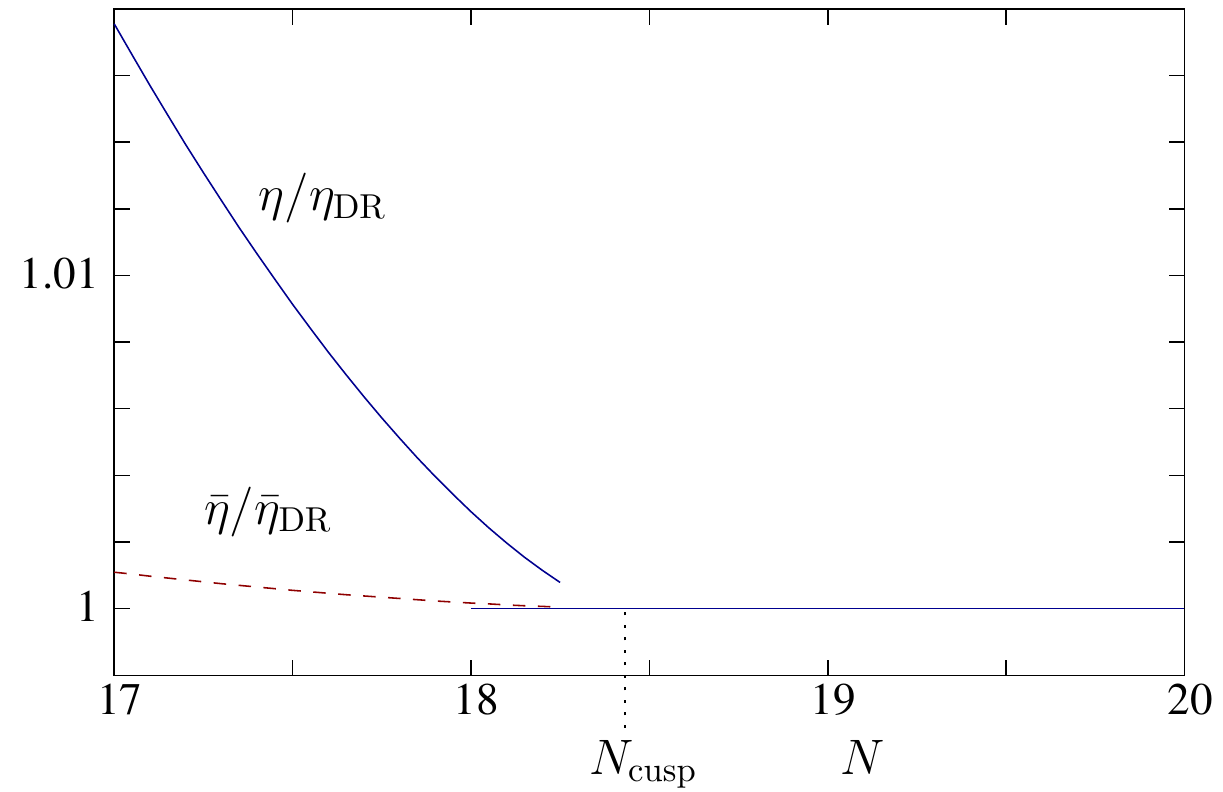}  
\caption{Anomalous dimensions $\eta$ (full line) and $\bar\eta$
  (dashed line), normalized by their dimensional-reduction value
  $\epsilon/(N-2)$, as a function of $N$. For $N>N_\cusp$, the stable
  fixed point leads to dimensional reduction. For $N_\DR<N<N_\cusp$
  we find a cuspless (unstable) fixed point which leads to dimensional
  reduction and a cuspy (stable) one associated with a breakdown of the
  dimensional reduction. For $N<N_\DR$, only the cuspy fixed point
  with dimensional-reduction breakdown remains.}
  \label{fig_etaetab}
\end{figure}

Once the cuspy fixed-point solution is (numerically) obtained, we can
derive the critical exponents, in particular the two anomalous
dimensions $\eta$ and $\bar\eta$. We focus here on the vicinity of
$N_\cusp$, which was not considered previously. We display in
Fig.~\ref{fig_etaetab} the two anomalous dimensions normalized by
their dimensional-reduction expression
$\eta_\DR=\etab_\DR=\epsilon/(N-2)$. The numerical determination of
the fixed-point solution near $N_\cusp$ is difficult because of the
presence of several fixed points which are close one to another, and
we were not able to determine the cuspy fixed point with sufficient
accuracy near $N_\cusp$. It is however clear numerically that for $N$
slightly larger than $N_\DR$ there exists indeed two fixed points,
with different anomalous dimensions.

We now discuss the behavior of the eigenfunction associated with the
cuspy perturbation around the cuspless fixed point. 
From the work of Sakamoto et al.,\cite{sakamoto06} one knows
that the physical eigenfunction $f_N(z)$ with a cusp is a
linear combination of two solutions of the eigenvalue equation,
$f_N^{(-)}(z)$ and $f_N^{(+)}(z)$, the former having a cusp when
$z\rightarrow 1$, \textit{i.e.}  $f_N^{(-)}(z)\simeq
\sqrt{1-z}\,[1+O(1-z)]$, and the latter one having only a subcusp,
\textit{i.e.} $f_N^{(+)}(z)\simeq (1-z)^{\alpha_{+}(N)}
[1+O(1-z)]$. Both functions, $f_N^{(-)}(z)$ and $f_N^{(+)}(z)$,
individually diverge in $z=-1$ and are therefore not acceptable
eigenfunctions. It is however possible to choose the coefficients of
the linear combination so that the divergence in $z=-1$ of the two
functions cancel. By continuity, we expect that this mechanism, which
has been checked to order $1/N^2$, still applies as one decreases $N$.

\begin{figure}[ht]
  \centering
  \includegraphics[width=\linewidth]{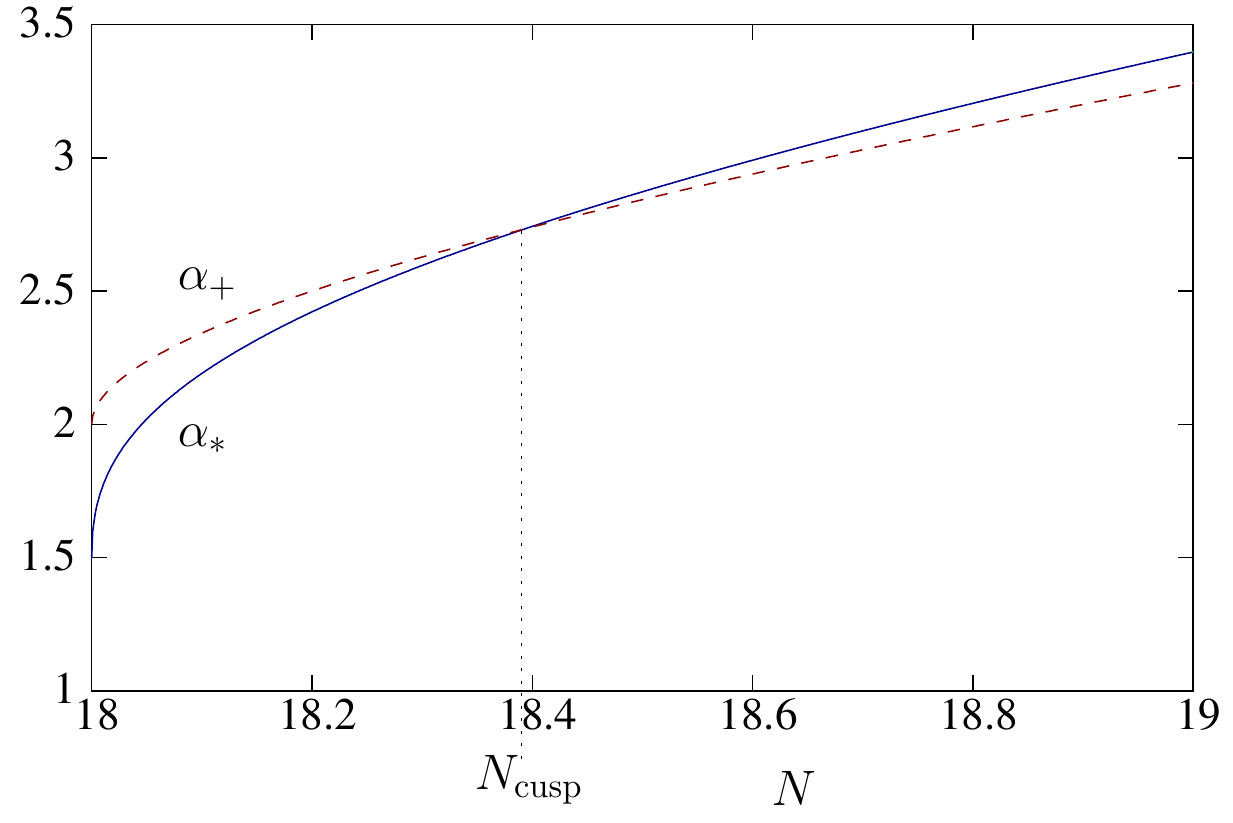}
  \caption{Behavior of the leading singularity $\alpha_*(N)$ of the cuspless critical fixed point (solid line) and of the sub-leading 
  singularity $\alpha_+(N)$ of the cuspy eigenfunction around this fixed point (dashed line).}
  \label{fig_alpha}
\end{figure}

One should therefore find two eigenfunctions, one with a cusp and one
with a subcusp in $(1-z)^{\alpha_{+}(N)}$, to ensure that a linear
combination of the two has a proper behavior in $z=-1$. The expression
for $\alpha_{+}(N)$ is obtained by imposing
\begin{equation}
  \label{eq_def_alpha+}
  \Lambda_{\alpha_++1}(\Delta_\star(1),\Delta'_\star(1))=\lambda(N)/\epsilon \,,
\end{equation}
where $ \Lambda_{\alpha_++1}$ is obtained from Eq.~(\ref{eq_Lambda}). We thus get
\begin{equation}
\label{eq_alpha+}
\alpha_{+}(N)=\frac 14\left(N-10+\sqrt{(N-2)(N-18)}\right) \,.
\end{equation}
We show in Fig.~\ref{fig_alpha} the behavior of $\alpha_+$ and $\alpha_*$ as a function of $N$. 
We observe that $\alpha_+$ is smaller than $\alpha_*$ for large $N$
(where $\alpha_+$ and $\alpha_*$ behave as $N/2-5+\mathcal O(1/N)$ and
$N/2-9/2+\mathcal O(1/N)$, respectively). For smaller values of $N$,
$\alpha_+$ becomes larger than $\alpha_*$ (in particular
$\alpha_*(N=18)=3/2$ and $\alpha_+(N=18)=2$). The two curves cross exactly
at $N_\cusp$. This is not a surprise since for $N=N_\cusp$,
$\lambda=0$, which implies that the conditions in Eqs.~(\ref{eq_def_alpha+})
and (\ref{eq_def_alpha*}) are degenerate.

The same analysis can be carried out at the two-loop order. To do so,
we have used the FRG equations derived in
Ref.~[\onlinecite{tissier06b}]. We can then obtain the eigenvalue
$\lambda(N)$ at order $\epsilon=d-4$, which allows us to determine
$N_{\cusp}$ at first order in $\epsilon$:
\begin{equation}
\label{eq_Ncusp}
N_{\cusp}(d)=2(4+3\sqrt 3) -3\frac{(2+3\sqrt{3})}{2}\epsilon \,.
\end{equation}
When compared to the result for $N_{\DR}$,
$N_{\DR}(d)=18-\frac{49}{5}\epsilon$,\cite{tissier06b} it can be seen that the absolute value
of the slope (with $\epsilon$ or $d$) is larger for $N_{\cusp}(d)$ than for $N_{\DR}(d)$. 
By extrapolating the results, we therefore find that the two lines
meet for $d=d_{\rm x}\simeq 4.4$ and $N=N_{\rm x}\simeq 14$. For the RFIM, where $N=1$ and $d_{\DR}\simeq 5.1$, one should 
thus expect another scenario for the destabilization of the dimensional-reduction fixed point than the one found near $d=4$. Indeed, 
for $d>4.4$, the cuspless critical fixed point disappears for $N_\DR$ when it is still stable with respect to a cuspy perturbation.

We would like to emphasize again that the annihilation and
disappearance of a pair of fixed points, with a square-root behavior
of a coupling constant like in Eq.~(\ref{eq_cuspless}), is a rather
common phenomenon in field theories, when several marginal operators
are compatible with the symmetries of the problem. This situation is
encountered in the Potts model,\cite{amit_potts, priest_potts} in
superconductors,\cite{lubensky,dasgupta} in Josephson junction
arrays,\cite{teitel} in He$_3$,\cite{jones,bailin} in smectic liquid
crystals,\cite{halperin} in electroweak phase transitions
\cite{lawrie,march-russel} and in frustrated magnets.\cite{diep} In
all of these cases, the two fixed points meet and annihilate at some
critical dimension. Beyond this dimension, the fixed-point
characteristics acquire an imaginary part and are no longer of
physical relevance. In the absence of a stable fixed point, the RG
flow typically leads the system toward a region where the potential is
unbounded from below because of operators of higher orders ($\phi^6$
terms for instance). This is in general interpreted as signaling the
occurrence of a first-order transition.

This is however {\it not} what we find in the numerical analysis of the
FRG flow equations for the RFIM. The typical situation is that there does
exist a fixed point beyond the line where the two cuspless fixed points
annihilate. In the next section, we present a toy model which we use
to illustrate how a cuspy fixed point can emerge continuously from the
annihilation of two cuspless fixed points. This unusual situation is made
possible because we are renormalizing a full function, while in the
situations previously mentioned, only a finite number of coupling constants are considered.

\section{Toy model}
\label{sec:toy_model}

We treat here a partial differential equation which is a
generalization of the 1-loop flow equation of the RF$O(N)$M in
Eq.~(\ref{eq_flow_delta}). We consider a function $\Delta(z)$ where
$z\in[-1,1]$. The evolution under the RG flow is given by the
following equation:
\begin{equation}
  \label{eq:betatoy}
  \begin{split}
  \partial_t
  \Delta_k(z)=&\Delta_k(z)-\Delta_k(z)\Delta'_k(z)-(\Delta_k(z)-z \Delta_k'(z)) \Delta_k(1)\\&
  +B\big [ \Delta_k(1)-z \Delta_k(z)\big ]\big [\Delta_k(z)+z \Delta_k'(z)\big ] +\\&
  \frac A2 (1-z^2)\Delta_k'(z)\big [2z\Delta_k'(z)-(1-z^2)\Delta_k''(z)\big ] \,.
  \end{split}
\end{equation}
The beta function depends on two parameters, $A$ and $B$, which
replace the two parameters $d$ and $N$ of the RF$O(N)$M.

We start our study of the toy model by a determination of the region of
parameters where cuspless fixed points exist. Assuming for now 
that the function $\Delta_k$ is sufficiently regular, ({\it i.e.}, that the
first derivative is finite in $z=1$), we get the following flow equations:
\begin{align}
  \partial_t \Delta_k(1)=&\Delta_k(1)\big [1-\Delta_k(1) \big ]\\
  \partial_t \Delta_k'(1)=&-B\big [\Delta_k(1)+\Delta_k'(1)\big ]^2\\\nonumber&
  +\Delta_k'(1)\big [1-(1+2A)\Delta_k'(1)\big ] \,.
\end{align}
Note that one has the property of the RF$O(N)$M that the flow of
$\Delta_k(1)$ depends on $\Delta_k(1)$ only and that the flow of
$\Delta_k'(1)$ depends on $\Delta_k(1)$ and $\Delta_k'(1)$ only.  The
``critical'' fixed-point solution of interest is $\Delta_\star(1)=1$,
which is once unstable in the direction $\Delta(1)$ (the associated
eigenvalue is negative). The beta function for $\Delta'(1)$, which is
a polynomial in $\Delta'(1)$, admits a real fixed-point solution for
\begin{equation}
  \label{eq:bcusp}
  B\leq B_\DR(A)=\frac1{8(1+A)} \,.
\end{equation}
If this condition is fulfilled, the solution reads
\begin{equation}
  \label{eq:deltap1pf}
  \Delta'_\star(1)= \frac{1-2B-\sqrt{1-8B(1+A)}}{2(1+2A+B)} \,.
\end{equation}
There is also a conjugate fixed point, with a plus sign in front of
the square root, which has at least two unstable directions and is
therefore not associated with a critical point.

\begin{figure}[tb]
  \centering
  \includegraphics[width=.9\linewidth]{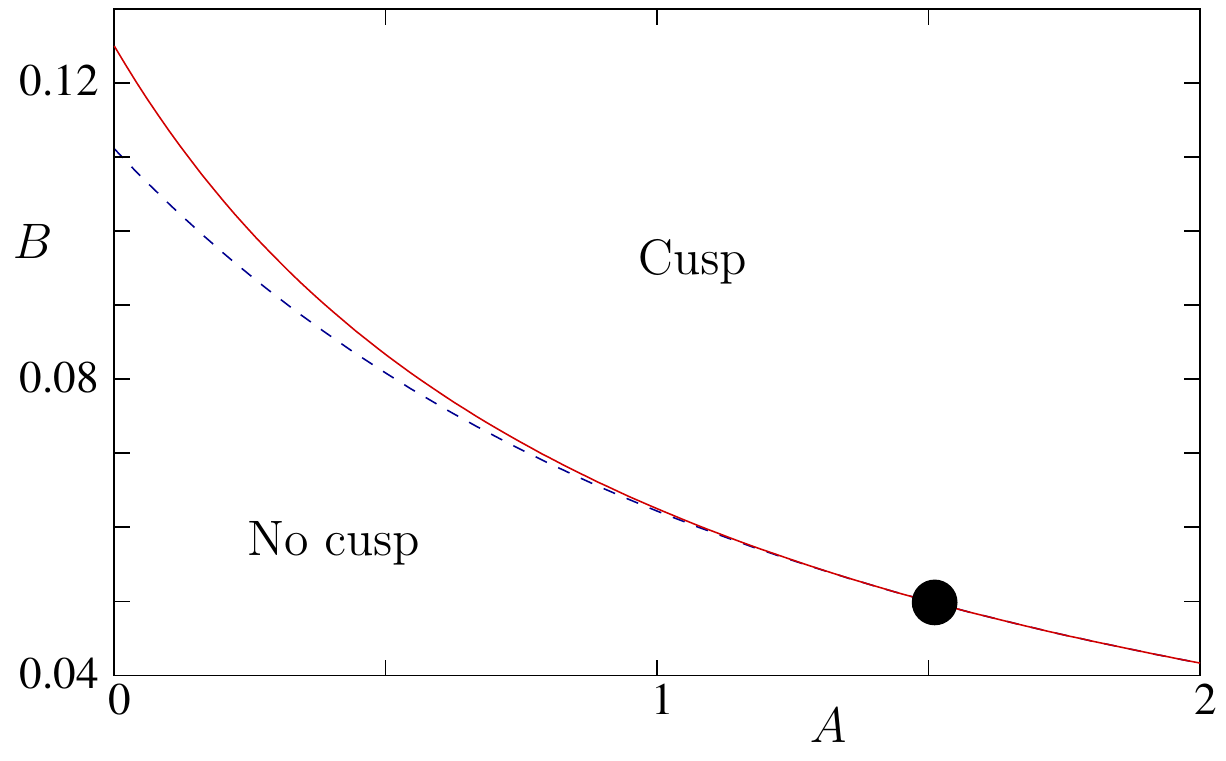}
  \caption{Toy model phase diagram in the $(A,B)$ plane. The curve
    $B_\dr(A)$ (full red line) is the boundary at which the cuspless
    fixed points present below it merge and annihilate. The curve
    $B_\cusp(A)$ (dashed blue line) is where the cuspless critical
    fixed point becomes unstable to a cuspy perturbation. The black
    point at $(A_{\rm x}=3/2, B_{\rm x}=1/20)$ is the intersection
    between the two curves. To the left of this point, when $B$ is
    increased, the cuspless fixed point first exchanges stability with
    a cuspy fixed point for $B=B_\cusp$, before it disappears at
    $B=B_\dr$. To the right of this point, the cuspless critical fixed
    point is stable up to $B=B_\dr$. Between the two lines, the
    cuspless critical fixed point exists but is unstable with respect
    to a cuspy perturbation. Note the similarity with
    Fig.~\ref{fig_phase_diagram}, except that the regions with or
    without cusp are not located in the same place.}
  \label{fig:AB}
\end{figure}

We now consider the vicinity of the cuspless critical fixed point and
derive the eigenvalue $\Lambda_{p+1}$ associated with a perturbation
whose functional form near $z=1$ starts with $\sim(1-z)^p$. A simple
calculation leads to
\begin{equation}
  \label{eq:Lambda}
  \begin{split}
\Lambda_{p+1}=&1+\Delta_\star(1)[p-1-B(p+1)]\\&-\Delta'_\star(1)(p+1)[1+2Ap+B] \,.
  \end{split}
\end{equation}
The eigenvalue $\lambda=\Lambda_{3/2}$ is of particular interest because it is associated with the cuspy direction. 
In the region $B<B_\dr$ where the cuspless critical fixed point exists we find that the
cuspy direction is a relevant perturbation around the latter for $B_\cusp(A)<B<B_\dr(A)$ and $A<A_{\rm x}=3/2$, with
\begin{equation}
\begin{aligned}
  \label{eq:Bcusp}
&B_\cusp(A)=\\&\frac{-16-21A-9A^2+(4+3A)\sqrt{25+18A+9A^2}}{36(1+A)} \,.
\end{aligned}
\end{equation}
Observe that $B_\cusp(A=A_{\rm x}=3/2)=B_\dr(A=A_{\rm x}=3/2)=1/20$ and that the eigenvalue $\lambda$ 
is then equal to zero, which means that, for this particular value of $A$, the cusp is marginal when the cuspless fixed
point vanishes. We summarize these findings  in Fig.~\ref{fig:AB}.

Another way of presenting the results is to evaluate $\lambda$ along the curve $B=B_\DR(A)$. 
We then find 
\begin{equation}
  \label{eq:lambcusp}
 \lambda(A,B_\dr(A))=\frac{-3+2 A}{4(3+4A)} \,.
\end{equation}
For $A<A_{\rm x}=3/2$, the cuspy direction is already relevant when
$B=B_\DR(A)$. This is the typical situation encountered close to $d=4$
in the RF$O(N)$M at one-loop order. On the contrary, for $A>A_{\rm
  x}=3/2$, the cuspy direction is still irrelevant when the cuspless
fixed points annihilate for $B=B_\DR(A)$, which is the typical
situation for the RFIM close to $d=5.1$.

We now study how a cuspy fixed point can appear when the cuspless
fixed points annihilate and disappear. We focus on the immediate
vicinity of $B_\dr$ and define, for a given $A$,
$B=B_\dr(A)+\epsilon'$. We anticipate that the cusp should appear in a
boundary layer around $z=1$ that shrinks to zero as $\epsilon'$ goes
to zero. We therefore make the following ansatz:
\begin{equation}
  \label{eq:def_f}
  \Delta_k(z)=1-\epsilon  f_k\left(\sqrt{\frac{1-z}\epsilon}\right) \,, 
\end{equation}
where $\epsilon \to 0$ when $\epsilon'\to 0$. After inserting this expression in the flow equation, Eq.~(\ref{eq:betatoy}), and 
expanding at leading order in $\epsilon$, we get
\begin{equation}
  \label{eq:flow_f}
\begin{split}
\partial_t f_k(y)=&-\frac 1{16(1+A)}\Bigg\{\frac{9+8A}yf_k'(y)[f_k(y)-f_k(0)]\\&+2y^2+2f_k(y)+(7+8A)[2f_k(0)-yf_k'(y)]
\\&+4A(1+A)f_k'(y)[f_k'(y)+yf_k''(y)]\big\}
  \end{split}
\end{equation}
where $y=\sqrt{(1-z)/\epsilon}$. Note that $\epsilon'$ which measures
the distance to $B_\DR$ does not appear in this equation. This means
that, at least at this level, we are unable to relate the
typical size of the boundary layer to the distance to $B_\DR$. We simply 
assume here that both tend to zero simultaneously.

We are interested in the fixed-point solution of the above flow
equation. For $y\gg 1$, {\it i.e.} outside the boundary layer, the
fixed-point function behaves as
\begin{equation}
  \label{eq:f_large_y}
  f_\star(y)\sim_{y\gg 1}\frac {y^2}{3+4A} \,.
\end{equation}
When inserted in Eq.~(\ref{eq:def_f}) this leads to $
\Delta_*(z)=1-(1-z)/(3+4A)$, which coincides with the expansion near
$z=1$ of the cuspless fixed point in $B=B_\DR(A)=1/[8(1+A)]$ [see
Eq.~(\ref{eq:deltap1pf})], \textit{i.e.} for $\epsilon'=0$.

Expanding now Eq.~(\ref{eq:flow_f}) for small $y$ (inside the boundary
layer, where $y \ll 1$), we find that the flow equation of
$f^{(p)}_k(0)$ depends only on the $p+1$ first derivatives at the
origin. We can therefore solve iteratively the fixed-point solution
and express the derivatives of $f_*(y)$ at the origin as a function of
one unknown, $f_\star(0)$. We find in particular that
\begin{align}
  \frac{f'_\star(0)^2}{f_\star(0)}&=-\frac{16(1+A)}{(3+2A)^2}\\
f''_\star(0)&=\frac{2(5+8A)}{3(9+16 A + 8 A^2)} \;.
\label{eq:f2}
\end{align}
This allows us to make predictions for the behavior of the original function. When $1-z\to 0$ (inside the boundary layer), we expand
$\Delta_\star(z)$ as
\begin{equation}
\Delta_\star(z)=\Delta_\star(1)-a_\star\sqrt{1-z}+\Delta_{\star,1}(1-z)+\cdots
\end{equation}
$\Delta_{\star,1}$ should not be interpreted here as the first derivative of  $\Delta_\star(z)$ in $z=1$ because of the singular dependence in 
$\sqrt{1-z}$. We then derive that
\begin{align}
  \frac{a_\star^2}{1-\Delta_\star(1)}&= \frac{f'_\star(0)^2}{f_\star(0)}=-\frac{16(1+A)}{(3+2A)^2}\\
\Delta_{\star,1}&=\frac {f''_\star(0)}2=\frac{5+8A}{3(9+16 A + 8 A^2)} \;.
\label{eq:delta2}
\end{align}
A direct comparison of Eq. (\ref{eq:delta2}), which is valid when
$B\to B_\dr(A)^+$, with the result for $B= B_\dr(A)$, obtained from
Eq.~(\ref{eq:deltap1pf}) with $B=B_\dr(A)=1/[8(1+A)]$ or from the
outer boundary-layer solution described above, shows that
$\Delta_{\star,1}$ is in general discontinuous for $B=B_\DR$, except
in $A=3/2$ where
$\Delta_{\star,1}\vert_{B_\dr^+}=\Delta_{\star,1}\vert_{B_\dr}=1/9$. On
the contrary, $\Delta_\star(1)$ is always continuous in $B_\dr$ and so
is the amplitude of the cusp $a_\star$ that continuously goes to zero
as $B\to B_\dr$. As a consequence, the critical exponents that depend
only on $\Delta_\star(1)$ and on the amplitude of the cusp, which is
the case of the exponent $\nu$ of the correlation length and of the
anomalous dimensions,\cite{tissier06b} are continuous. On the other
hand, eigenvalues that depend on $\Delta_{\star,1}$ are not: this is
the case for instance of the eigenvalue $\lambda$ associated with a
cuspy perturbation.  
Although $\Delta_\star(z)$ is a continuous
function of $B$ at fixed $z$ when B increases from $B_\DR$, some
properties of the fixed point may be discontinuous, which is a very
unusual situation in the RG of critical phenomena.

\begin{figure}[tb]
  \centering
  \includegraphics[width=.9\linewidth]{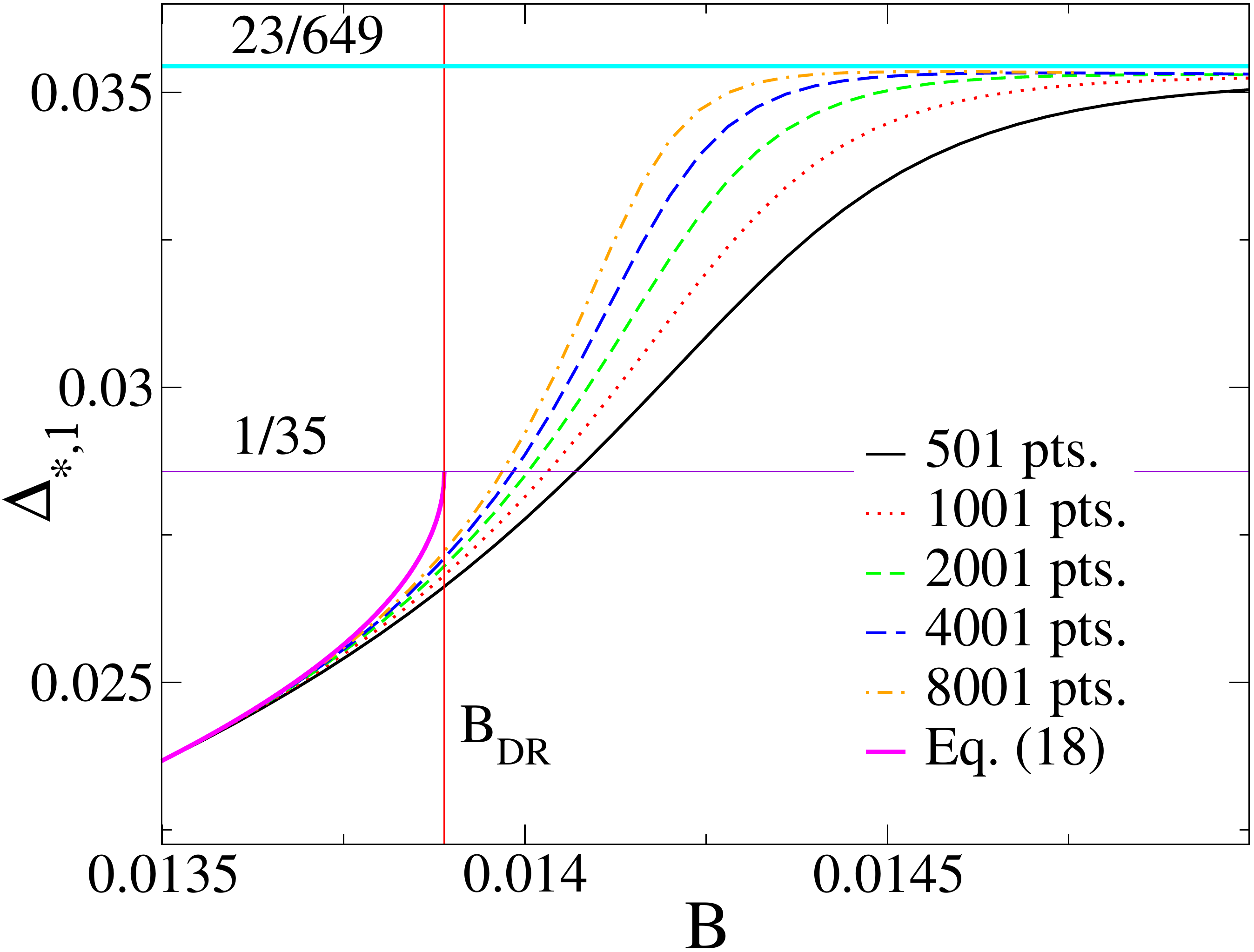}
  \caption{ $\Delta_{\star,1}$ as a function of $B$ for
    $A=8$. For $B<B_\DR(A=8)=1/72$, the numerical solution approaches
    the predicted result in the absence of a cusp, $1/35$, when finer
    meshes (larger number of points) are considered. For $B$
    approaching $1/72$ from above, the numerical solution tend to
    $23/649$, which is the solution extracted from the analysis of the
    boundary layer, see Eq.~(\ref{eq:delta2}).  }
  \label{fig:bl}
\end{figure}

\begin{figure}[htbp]
  \centering
  \includegraphics[width=.9\linewidth]{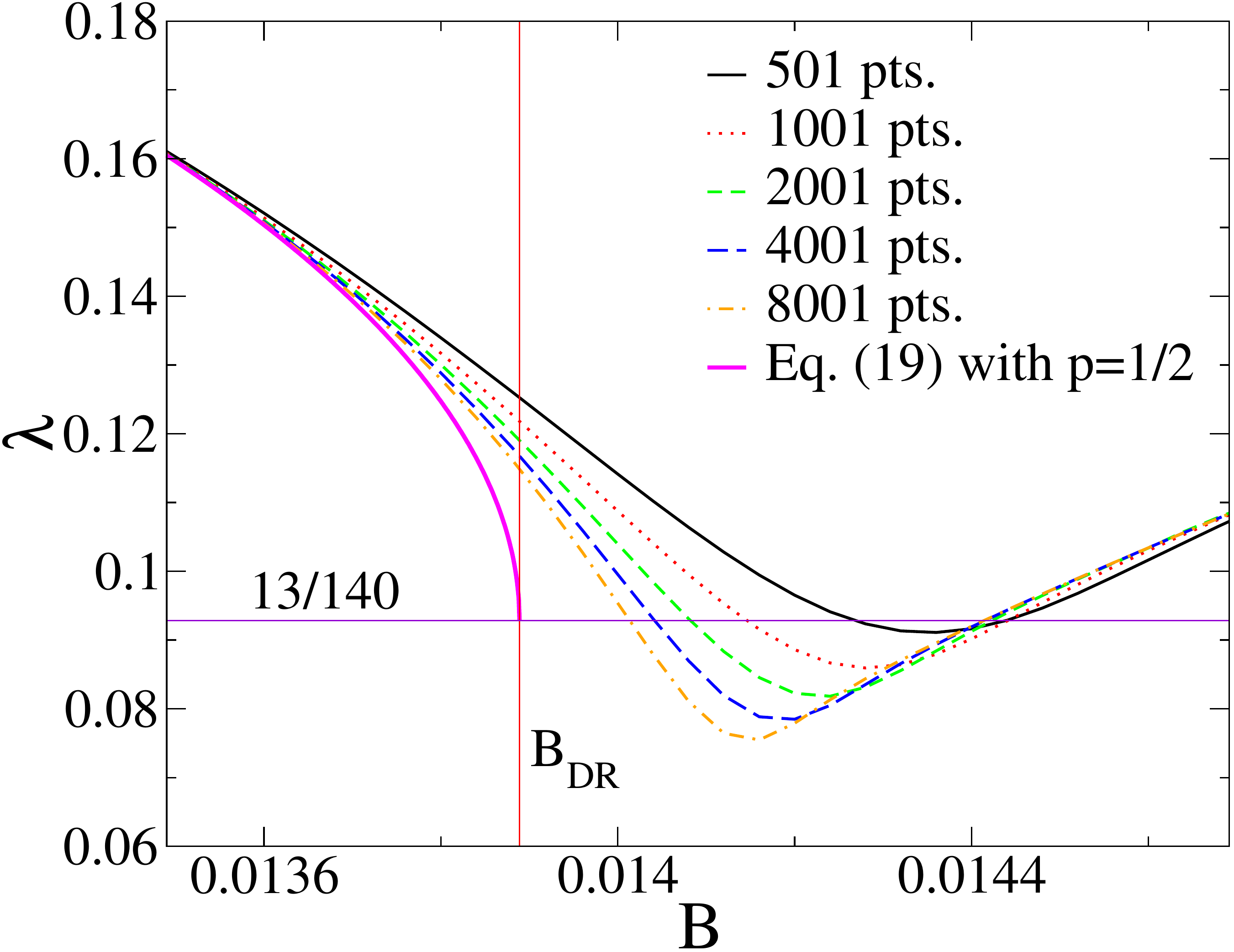}
  \caption{Eigenvalue $\lambda$ of the cuspy direction around the
    stable fixed point as a function of B around $B_\DR$ for
    $A=8$. For $B<B_\DR$, the numerical results tend to the exact
    analytical solution. For larger $B$, the numerical solution
    clearly tends to a lower value for finer meshes (larger number of
    points). This indicates that the eigenvalue $\lambda$ is
    discontinuous at $B_\DR$.}
  \label{fig:cuspy_dr}
\end{figure}

We have checked by a direct numerical integration of the flow equation
in Eq.~(\ref{eq:betatoy}) that the behaviors predicted above are
indeed observed. This is illustrated for $A=8>A_{\rm x}$:
Fig.~\ref{fig:bl} for $\Delta_{\star,1}$ and Fig.~\ref{fig:cuspy_dr}
for the eigenvalue $\lambda$.


Finally, in the particular case $A=0$, we have been able to solve analytically the
fixed-point equation in Eq. (\ref{eq:f_large_y}). There is a unique family of solutions parametrized by $f_*(0)$ . It can be
expressed in terms of the Lambert function $W(x)$ (solution of $W\exp W=x$):
\begin{equation}
\begin{aligned}
  \label{eq:soltoy}
& f_*(y)=\\&
\frac {y^2} 3+f_*(0)\left\{-1+2\exp\left[1+W\left(-\frac{2y^2+9 f_*(0)}{9ef_*(0)}\right)\right]\right\} \,.
\end{aligned}
\end{equation}
It is easily checked that the above solution satisfies the limiting
behaviors described above when $y\to \infty$ and $y \to 0$. Note that
this case corresponds to the region where $A<A_{\rm x}$ and that the
above cuspy fixed point which emerges from the merged cuspless ones is
unstable. Another cuspy fixed is present and is stable for $B>B_\cusp$
[here, $B_\cusp(A=0)=1/9$]. The generic situation is that there are
two cuspy fixed points above $B_\dr$, one stable and one unstable, one
that appears through a boundary-layer mechanism at $B_\dr$ and one
that is already present below $B_\dr$.

\section{The short- and long-range RFIM}
\label{sec:RFIM}

We have next investigated the $d$-dimensional RFIM ($N=1$). In this
case however, a nonperturbative FRG (NP-FRG) is
required.\cite{tarjus04,tissier06,tissier11} The central quantity is
now the dimensionless cumulant of the renormalized random field
$\delta_k(\varphi_1,\varphi_2)$. It follows an FRG equation that is
coupled to other functions describing the flow of the
disorder-averaged effective action (the latter is described in a
derivative expansion\cite{tarjus04,tissier11}).  The flow equations
can be symbolically written as
\begin{equation}
\label{eq_flow_dimensionless}
\begin{split}
&-\partial_t u'_k(\varphi)=\beta_{u'}(\varphi),\\&
-\partial_t z_k(\varphi)=\beta_{z}(\varphi),\\&
-\partial_t\delta_k(\varphi_1,\varphi_2)=\beta_{\delta}(\varphi_1,\varphi_2),
\end{split}
\end{equation}
where as before $t=\log(\Lambda/k)$; $u_k(\varphi)$ is the dimensionless effective average potential (\textit{i.e.}, 
the local component of the disorder-averaged effective action) and $z_k(\varphi)$ is the dimensionless 
function describing the renormalization of the field. The beta functions themselves depend on $u_k'$, $z_k$, $\delta_k$ and on their derivatives. 
Their expressions are given in Ref.~[\onlinecite{tissier11}]  and are not reproduced here.

We consider first the usual short-range RFIM in which both the
interactions and the random-field correlations are short-ranged [see
Eq. (\ref{eq_ham_rfon}) and below]. Fixed points are studied by
setting the left-hand sides of the equations in
Eq. (\ref{eq_flow_dimensionless}) to zero. The zero-temperature fixed
point controlling the critical behavior has been determined in a
previous investigation:\cite{tissier11} above a dimension $d_{\DR}$
close to $5.1$, there exists a cuspless fixed point that can be
reached when starting from a regular, cuspless, initial condition. The
presence or absence of a cusp now refers to the dependence of
$\delta_k(\varphi_1,\varphi_2)$ on the field difference $\varphi_2 -
\varphi_1$. For its description, it turns out to be more convenient to
change variable from $\varphi_1$ and $\varphi_2$ to
$\varphi=(\varphi_1+\varphi_2)/2$ and $y=(\varphi_1-\varphi_2)/2$. The
putative cusp is now in the variable $y$. For $d>d_{\DR}$, the
(critical) cuspless fixed point, which is characterized in the limit
$y \rightarrow 0$ by
\begin{equation}
\label{eq_cuspless_delta}
\delta_*(\varphi,y) = \delta_{*,0}(\varphi) + \frac{1}{2} \delta_{*,2}(\varphi)y^2  + \mathcal O(\vert y \vert^3)\, ,
\end{equation}
is stable with respect to cuspless perturbations, except of course for
the relevant direction that corresponds to a fine-tuning to the
critical point. As already stressed, such a fixed point corresponds to
the $d\rightarrow d-2$ dimensional reduction.

We have also investigated the stability of the cuspless,
dimensional-reduction, fixed point with respect to a cuspy
perturbation. We have followed the procedure described above for the
RF$O(N)$M near $d=4$. We search for a physical eigenfunction
$f(\varphi,y)$ with a linear cusp in $y$ that is a linear combination
of two solutions of the associated eigenvalue equation,
$f^{(-)}(\varphi,y)$ and $f^{(+)}(\varphi,y)$, the former having a
cusp when $y\rightarrow 0$, \textit{i.e.} $f^{(-)}(\varphi,y)\simeq
\vert y\vert [f_-(\varphi)+O(y^2)]$, and the latter having a subcusp
only, \textit{i.e.} $f^{(+)}(\varphi,y))\simeq \vert
y\vert^{\alpha_{+}(d)} [f_{+}(\varphi)+O(y^2)]$ with $\alpha_{+}(d)$
odd or noninteger. The linear combination should ensure that all
divergences are cancelled and that the physical eigenfunction is
defined for all values of $y$.

The corresponding eigenvalue $\lambda$ can then be determined by considering the vicinity of the fixed point with  
$\delta_k(\varphi,y)\simeq \delta_*(\varphi,y) +k^{\lambda} f(\varphi,y)$ and  $f(\varphi,y) \simeq \vert  y \vert f_-(\varphi)$ 
when $y \rightarrow 0$. By linearizing the flow equation for $\delta_k$ around $\delta_*$, fixing $u'_k(\varphi)$ 
and $z_k(\varphi)$ to their fixed-point values, and expanding around 
$y= 0$, it  is easy to derive that $ f_-(\varphi)$ satisfies the following eigenvalue equation:
\begin{equation}
 \label{eq_lambda}
\begin{split}
&\lambda f_-(\varphi) = \frac{1}{2} (d - 4  +3 \eta ) f_-(\varphi)  +  \frac{1}{2}  (d -4 + \eta ) \varphi f_-'(\varphi) 
+ \\& v_{d}\,   \tilde{\partial}_{t} \int^{\infty}_{0} dx \: x^{\frac{d}{2}-1}  
\bigg\lbrace \frac{3}{2}  f_-(\varphi) \Big(  4 z_{*}'(\varphi) p_{*}(x,\varphi) p^{(0,1)}_{*}(x,\varphi) +\\&
 4 [z_{*}(\varphi) + s' (x)]  p^{(0,1)}_{*}(x,\varphi)^{2}+[z_{*}''(\varphi) - \delta_{*, 2} (\varphi)]  p_{*}(x,\varphi)^{2}  \Big) 
\\& +  3     f_-'(\varphi)  p_{*}(x,\varphi)  \Big( 2[z_{*}(\varphi) + s' (x)]  p^{(0,1)}_{*}(x,\varphi)  
+ \\& z_{*}'(\varphi) p_{*}(x,\varphi) \Big)+ f_-''(\varphi)  [z_{*}(\varphi)+ s' (x)] p_{*}(x,\varphi)^2   \bigg\rbrace   \,,
\end{split}
\end{equation}
where $v_d^{-1}=2^{d+1}\pi^{d/2} \Gamma(d/2)$, partial derivatives are
denoted by superscripts in parentheses, and $x$ is the square of the
dimensionless momentum; $p_{*}(x,\varphi ) = [ x z_{*}(\varphi) + s(x)
+ u_{*}''(\varphi) ]^{-1}$ is the (dimensionless) ``propagator'',
\textit{i.e.} the so-called ``connected'' $2$-point correlation
function, and $s(x)$ is a (dimensionless) cutoff function.  (Choices
of appropriate functional forms for $s(x)$ are discussed in
Ref.~[\onlinecite{tissier11}].) Finally, $\widetilde \partial_t$ is an
operator acting only on the cutoff function $s(x)$ (appearing
explicitly or through the dimensionless propagator) with
$\widetilde \partial_t s(x) \equiv (2-\eta)s(x)-2xs'(x)$.  In deriving
the above equation, we have used the fact that $\bar\eta=\eta$ and
$\delta_{*,0}(\varphi)=z_{*}(\varphi)$, which are properties of the
cuspless, dimensional-reduction, fixed point resulting from the
underlying supersymmetry.\cite{tissier11}

An equation for the fixed-point function $\delta_{*, 2} (\varphi)$
that appears in Eq.~(\ref{eq_lambda}) can be also derived by inserting
the expansion in powers of $y$ of $\delta_*(\varphi,y)$ [see
Eq.~(\ref{eq_cuspless_delta})] in the corresponding beta function in
Eq.~(\ref{eq_flow_dimensionless}).  The algebra is straightforward but
cumbersome and leads to:
\begin{equation}
 \label{eq_delta2}
\begin{split}
&0= (d - 4 +2 \eta_{*} )\delta_{*, 2}( \varphi ) +  \frac{1}{2}  (d -4 + \bar{\eta}_{*} ) \varphi\delta_{*, 2}'( \varphi ) + \\&
v_{d}   \tilde{\partial}_{t} \int^{\infty}_{0} dx \: x^{\frac{d}{2}-1}  \Bigg\lbrace \bigg( \, 4   
\tilde{p}^{(0,1)}_{*}(x, \varphi )  \tilde{p}^{(0,2)}_{*}(x, \varphi )z_*'( \varphi ) \big [z_*(\varphi)  \\&
+ s'(x)\big] + 5  \tilde{p}^{(0,1)}_{*}(x, \varphi )^2 z_*'( \varphi )^2 +  \tilde{p}_{*}(x, \varphi )  
\big(2  \tilde{p}^{(0,2)}_{*}(x, \varphi ) \times \\&
z_*''( \varphi ) 
\big[z_*( \varphi ) + s'(x)\big] + 7  \tilde{p}^{(0,2)}_{*}(x, \varphi ) z_*'( \varphi )^2 +
 \\&4z_*'(\varphi ) \big(2z_*''( \varphi )   \tilde{p}^{(0,1)}_{*}(x, \varphi ) +  \tilde{p}^{(0,3)}_{*}(x, \varphi ) \big[\delta_{*,0}( \varphi ) 
+ s'(x)\big]\big) \big)\\
&- 2 \big[  \tilde{p}^{(0,2)}_{*}(x, \varphi )^2 -  \tilde{p}^{(0,1)}_{*}(x, \varphi ) \tilde{p}^{(0,3)}_{k}(x, \varphi) \big] \big[z_*( \varphi ) 
+ s'(x)\big]^2\\&
+ \frac{1}{2}    \tilde{p}_{*}(x, \varphi )^2 \big[z_*''( \varphi )^2+2z_*^{(3)}( \varphi )z_*'( \varphi )\big]
+  3 \,\delta_{*, 2}(\varphi)\times \\&
 \bigg(4  \tilde{p}^{(0,1)}_{*}(x, \varphi )  \tilde{p}_{*}(x, \varphi ) z_*'( \varphi )+ 
4  \tilde{p}^{(0,1)}_{*}(x, \varphi )^2 \big[z_*( \varphi) + s'(x)\big]\\&
+ \tilde{p}_{*}(x, \varphi )^2 z_*''( \varphi ) \bigg) + \delta_{*, 2}'( \varphi ) \bigg( 4\tilde{p}_{*}(x, \varphi) 
\big(2 \tilde{p}^{(0,1)}_{*}(x, \varphi ) \times \\&
\big[z_*( \varphi ) + s'(x)\big] +  \tilde{p}_{*}(x, \varphi )z_*'( \varphi ) \big)\bigg)+\delta_{*, 2}''( \varphi) 
\tilde{p}_{*}(x, \varphi )^2  \times \\&
\big[z_*( \varphi ) + s'(x)\big]  - \frac{3}{2} \delta_{*, 2}( \varphi )^2 \tilde{p}_{*}(x, \varphi )^2 \Bigg\rbrace \,,
\end{split}
\end{equation}
where we have used that $\delta_{*, 0} (\varphi)=z_{*}(\varphi)$. From
the knowledge of $u'_{*}(\varphi)$ and $z_{*}(\varphi)$, which are
obtained from two coupled equations (see
Ref.~[\onlinecite{tissier11}]), we first solve the equation for
$\delta_{*, 2} (\varphi)$ and then use the input to solve
Eq.~(\ref{eq_lambda}). All partial differential equations are
numerically integrated on a one-dimensional grid by discretizing the
field $\varphi$.

\begin{figure}[tbp]
\centering
\includegraphics[width=.9 \linewidth]{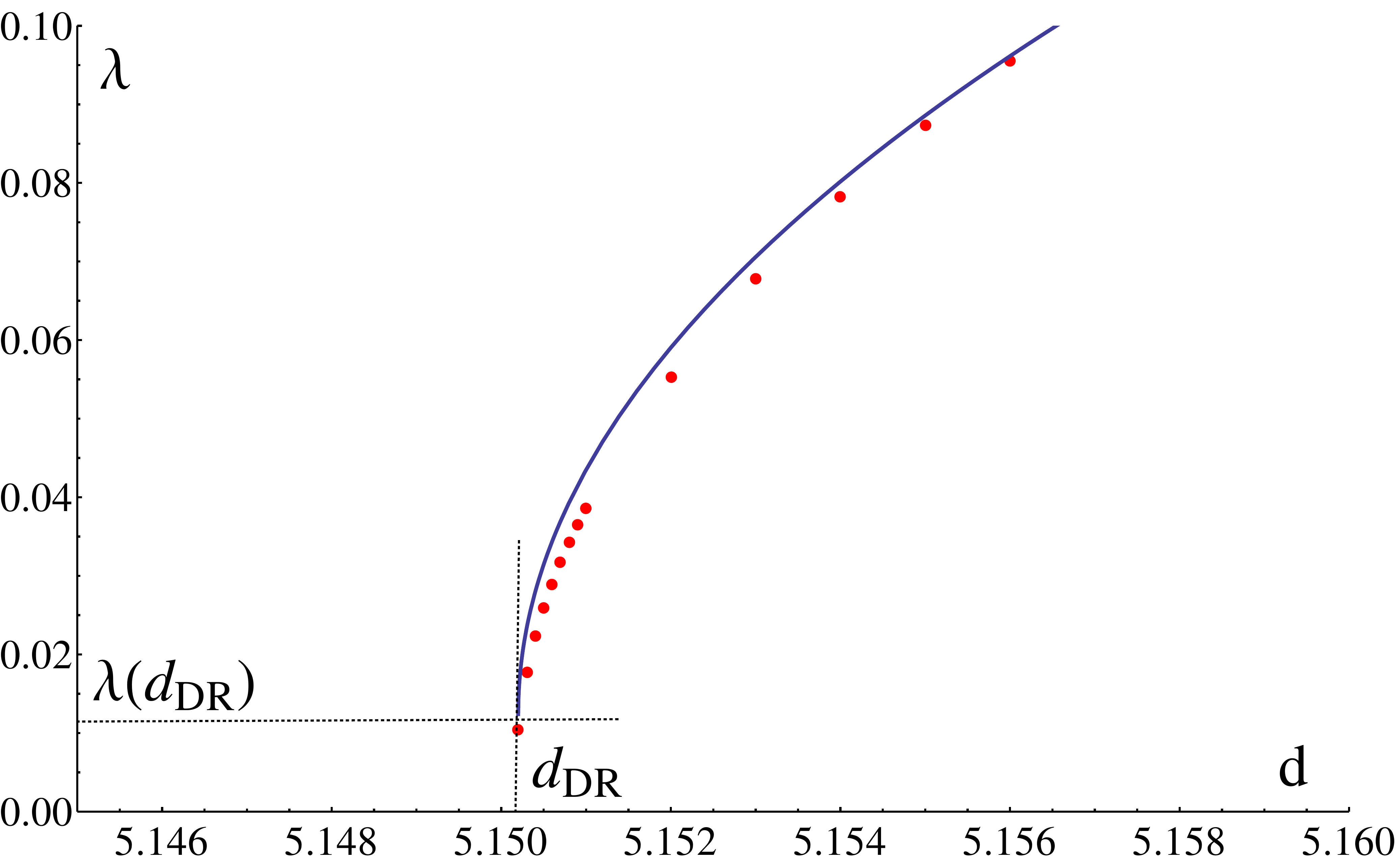}
\includegraphics[width=.9 \linewidth]{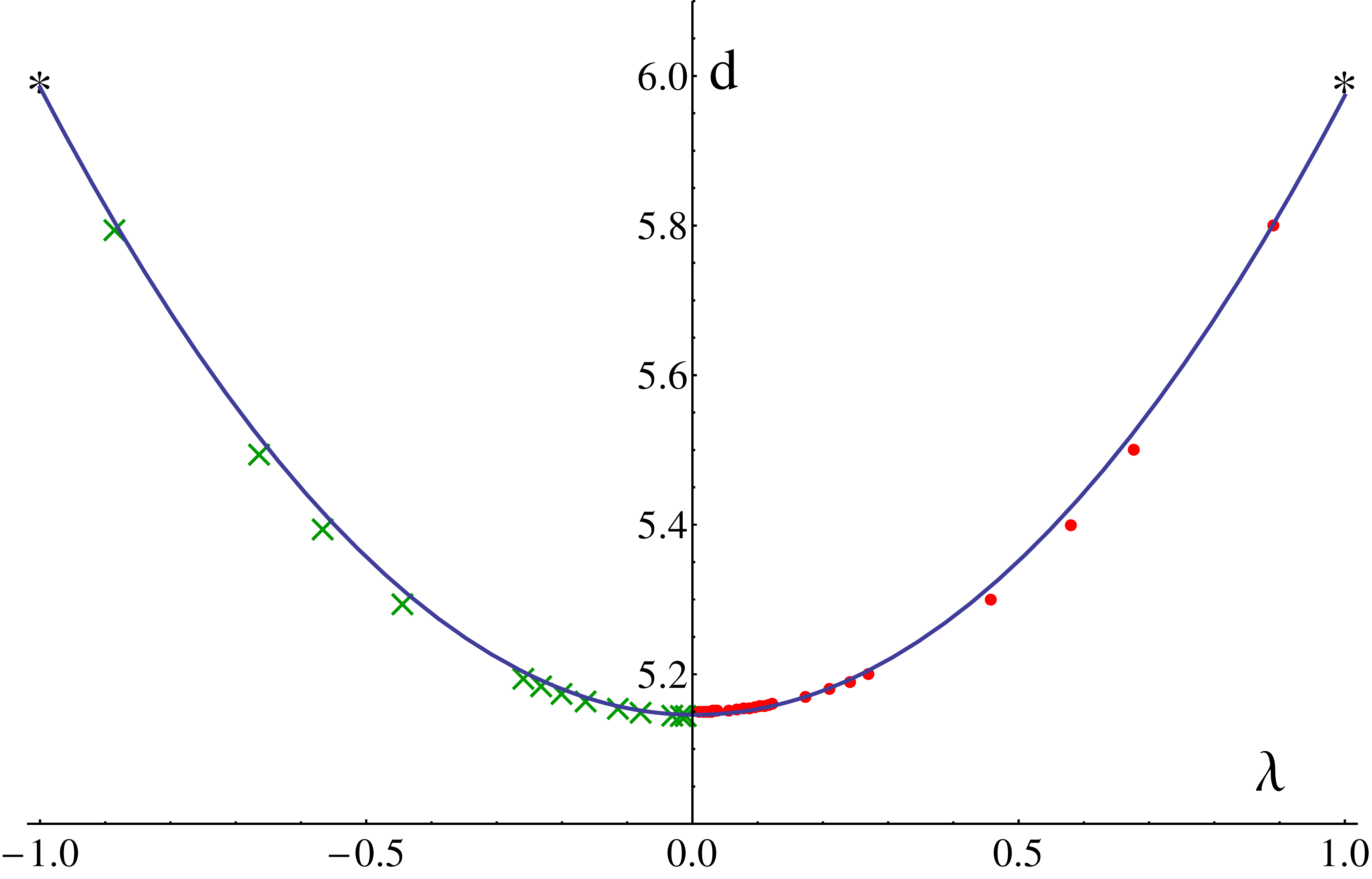}
\caption{ Short-range RFIM within the NP-FRG. Top figure: Variation
  with $d$ of the eigenvalue $\lambda$ associated with a cuspy
  perturbation around the stable cuspless fixed points when $d\geq
  d_{\DR}\simeq 5.1$. The dots (color online red) are obtained from
  the NP-FRG flow equations. The (color online blue) curve is a fit
  described in the text. At the dimension $d_\DR$, the eigenvalue is
  small but strictly positive. Bottom figure: Eigenvalue $\lambda$ of
  the cuspy perturbation around the stable cuspless fixed point (red
  points) and the unstable cuspless fixed point (green crosses). The
  black stars correspond to the exact values (-1 and +1) obtained in
  $d=6$. }
\label{fig_SR_RFIM}
\end{figure}

The resulting eigenvalue $\lambda(d)$ is plotted in
Fig.~\ref{fig_SR_RFIM}. Note that $\lambda$ can be calculated exactly
at the Gaussian fixed point that controls the critical behavior at and
above the upper critical dimension $d=6$, and one finds $\lambda=1$
with an associated eigenfunction $f(\varphi,y)=\vert y\vert$. As seen
in the figure, $\lambda$ is small but strictly positive when
$d=d_{\DR}$, in agreement with the phase diagram displayed in
Fig.~\ref{fig_phase_diagram}.

We have also checked that there is an additional solution
$f^{(+)}(\varphi,y)$ that is associated with the same eigenvalue
$\lambda$ and whose dependence on $y$ starts with a subcusp when
$y\rightarrow 0$. In addition, we have repeated the analysis for the
cuspless unstable fixed point that is conjugate to the above critical
one: it is characterized by the same $u'_{*}(\varphi)$ and
$z_{*}(\varphi)=\delta_{*, 0} (\varphi)$ but corresponds to another
solution $\delta_{*, 2} (\varphi)$ of Eq.~(\ref{eq_delta2}). The
eigenvalue associated with a cuspy perturbation around this fixed
point is plotted in the bottom panel of Fig.~\ref{fig_SR_RFIM} and it
merges with that for the other fixed point for $d=d_{\DR}$.  The cuspy
eigenvalues for both cuspless fixed points have a square root
behavior, as shown in the figure. We can fit these curves by a
parabola, $d(\lambda)=5.1503 - 0.0199\lambda + 0.8279\lambda^2$. We
observe that $\lambda(d_\dr)$ is slightly positive, as already
mentioned, and that a cuspy perturbation around the \textit{unstable}
cuspless fixed point is marginal in a dimension slightly larger than
$d_\DR$. From the results of the preceding sections, this indicates
that, in the case of the short-range RFIM, the breaking of dimensional
reduction is associated with the appearance of a cuspy fixed point
through a boundary layer. However, since $\lambda(d_\dr)$ is very
small, we expect that the unusual features that signals the presence
of a boundary-layer mechanism [in particular the discontinuity of the
coefficient $\delta_{*,2}(\varphi)$ of the term in $y^2/2$ of the
small $y$ expansion of $\delta_{*}(\varphi,y)$] to be almost
unobservable.

Consequently, we have also investigated the RFIM in the presence of
both long-range interactions, which decay in space as $\vert
x-y\vert^{-(d+\sigma)}$, and long-range correlations of the random
field that vary as $\vert x-y\vert^{-(d-\rho)}$. We have recently
shown that for a specific choice of the exponents characterizing these
long-range spatial dependences, namely $\rho=2-\sigma$, a
supersymmetry can still be present in the associated superfield
theory.\cite{baczyk13} This supersymmetry leads to a $d\rightarrow
d-2$ dimensional-reduction property. The corresponding cuspless fixed
point exists below a critical value $\sigma_{\DR}$, which in $d=3$ is
found between $0.71$ and $0.72$ depending on the precise choice of the
dimensionless cutoff function $s(x)$, and disappears above. (In this
case, the analogs of the lower and upper critical dimensions are a
critical value $\sigma=1$ above which there is no transition and a
critical value $\sigma=1/2$ below which the exponents are described by
mean-field theory.)

\begin{figure}[tb]
  \centering
  \includegraphics[width=\linewidth]{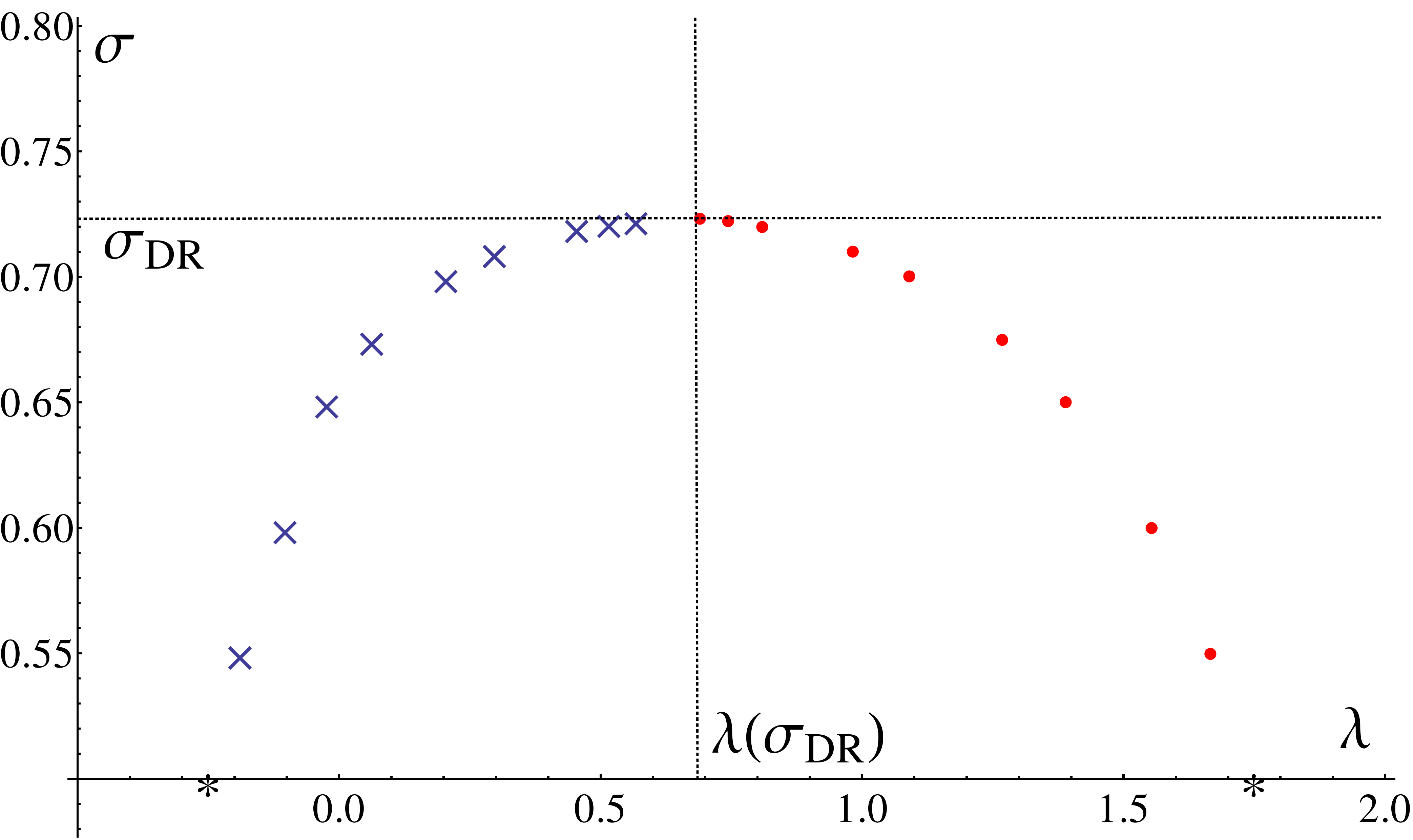}
  \caption{RFIM with long-ranged interactions and disorder correlation
    in $d=3$ within the NP-FRG. Variation of the eigenvalue $\lambda$
    associated with a cuspy perturbation around the stable cuspless
    fixed point (red dots) and around the unstable cuspless fixed
    point (blue crosses).  Note that $\lambda(\sigma_\dr)\simeq 0.7$
    is unambiguously strictly positive. The black stars correspond to
    the exact solutions (-1/4 and 7/4) at $\sigma=1/2$.}
  \label{fig_lr_ev}
\end{figure}

In this long-range model, we have repeated the analysis of a cuspy
perturbation around the stable and unstable cuspless fixed points. For
the stable (critical) fixed point the eigenvalue $\lambda$ decreases
from 7/4 for $\sigma=1/2$ to $\sim 0.7$ for $\sigma_{\DR}$. The
eigenvalue of the cuspy perturbation around the unstable fixed point
increases from -1/4 for $\sigma=1/2$ to $0.7$ for $\sigma_{\DR}$ and
passes through zero for $\sigma\simeq 0.66$. This is displayed in
Fig.~\ref{fig_lr_ev}. In this case, the eigenvalue associated with the
cuspy perturbation is unambiguously strictly positive when the two
cuspless fixed points coalesce, as seen in Fig.~\ref{fig_lr_ev}, and
the cuspy perturbation aound the unstable fixed point becomes marginal
for $\sigma_\cusp\simeq 0.65$, which is significantly different from
$\sigma_\DR$. We therefore expect a cuspy fixed point to appear
through a boundary layer for $\sigma>\sigma_\DR$ with a sizable
discontinuity in $\delta_{*,2}(\varphi)$ in $\sigma_\DR$. (Note that
$\delta_{*,2}(\varphi)$ is obtained as the second derivative of
$\delta_{*}(\varphi,y)$ with respect to $y$ in $y=0$ only in the
absence of a cusp.)

\begin{figure}[tb]
  \centering
  \includegraphics[width=.9\linewidth]{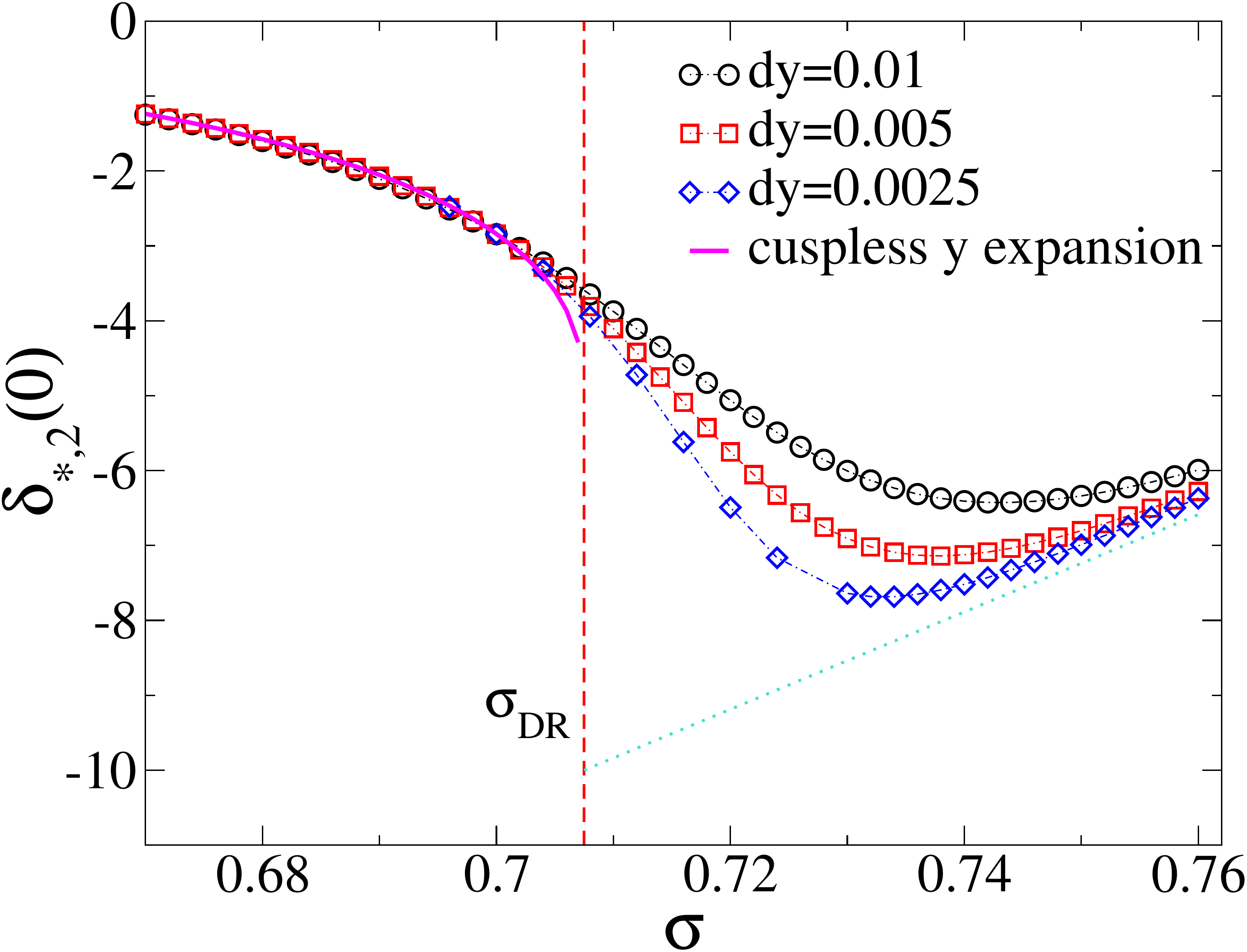}
  \includegraphics[width=.9\linewidth]{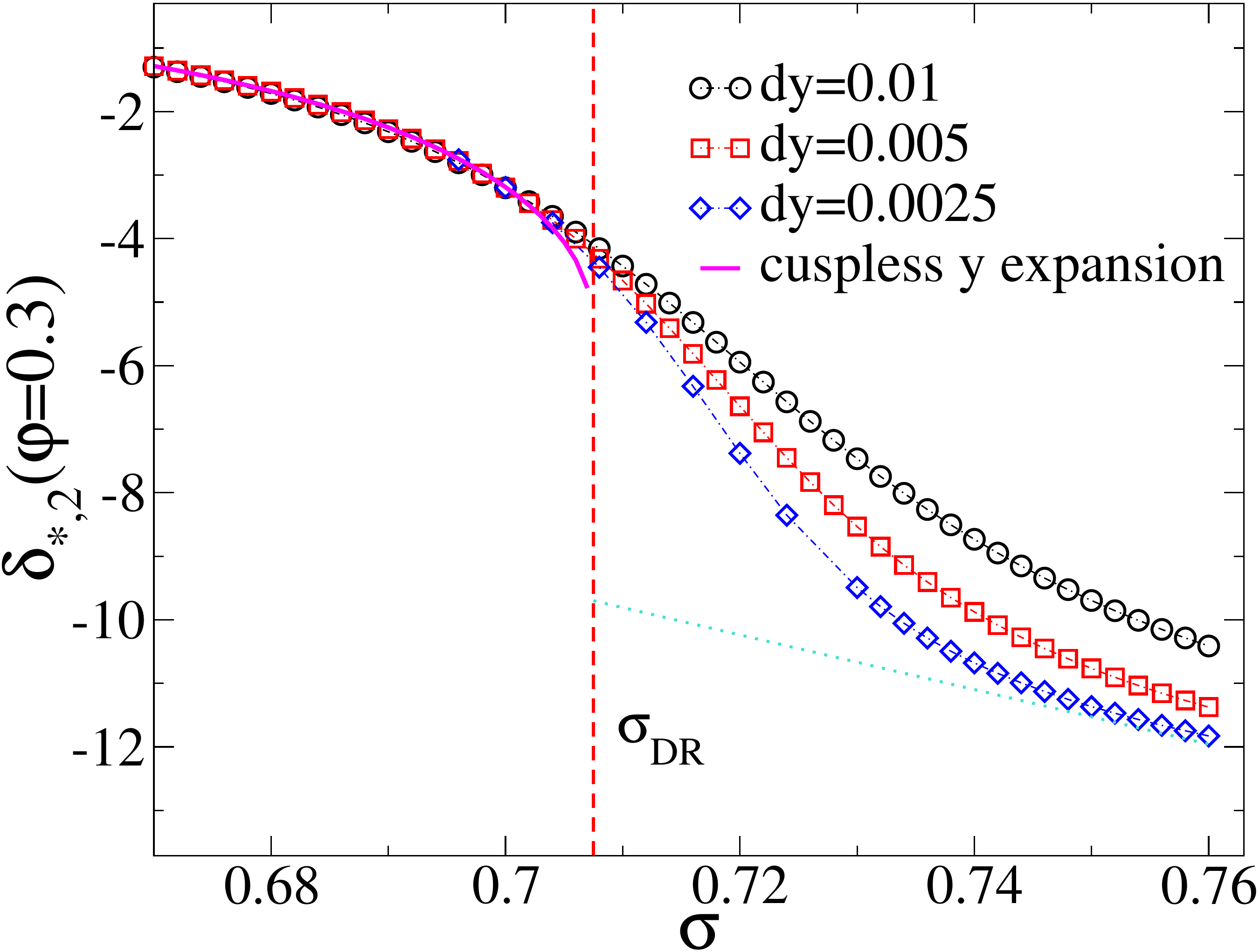}
  \caption{NP-FRG fixed-point solutions for the long-range RFIM in
    $d=3$. The coefficient $\delta_{*,2}(\varphi)$ of the $y^2/2$ term
    in the small $y$ expansion of $\delta_*(\varphi,y)$ evaluated in
    $\varphi=0$ (top panel) and in $\varphi=0.3$ (bottom panel) as a
    function of $\sigma$ around $\sigma_\DR\simeq 0.71$. When finer
    meshes (smaller $dy$) are considered, this quantity builds up a
    discontinuity, typical of the behavior expected from a
    boundary-layer mechanism. }
  \label{fig:bl_lr}
\end{figure}

To complement the above study of the stability of the cuspless fixed
points above $\sigma_{\DR}$, we have integrated the flow equations
[see eqs. (\ref{eq_flow_dimensionless})] without expanding in the
$y$-direction. We focus on the long-range model. We find strong
evidence for the occurrence of a boundary-layer mechanism for the
appearance of a stable cuspy fixed point above $\sigma_\DR$. This is
illustrated in Fig.~\ref{fig:bl_lr} where we plot
$\delta_{*,2}(\varphi)$ for two different values of $\varphi$ as a
function of $\sigma$ around $\sigma_\DR$. It can be seen that a
discontinuity builds up as the mesh size is decreased, very much as in
the toy model (see in particular Fig.~\ref{fig:bl}).

\section{Conclusion}

We have analyzed the mechanism by which the dimensional-reduction
result breaks down in the RF$O(N)$M by following the appearance,
disappearance, and change of stability of the (zero-temperature) fixed
points. We have combined the perturbative FRG results near $d=4$, at
one and two loops, the nonperturbative FRG results, in particular for
the short- and long-range RFIM, and a toy model. Dimensional reduction
for the critical behavior of the model is associated with a cuspless
fixed point and breaking of dimensional reduction with a cuspy fixed
point. (We recall that the cuspless or cuspy character refers to the
functional dependence of the dimensionless second cumulant of the
renormalized random
field\cite{tarjus04,tissier06,tissier06b,tissier11,baczyk13} and is
physically associated with the subdominant or dominant role of the
avalanches in the correlation functions.\cite{tarjus13})

The outcome of our study is an intricate scenario which is illustrated
in the $(N,d)$ phase diagram of Fig. \ref{fig_phase_diagram}.  There
are two different regimes separated by a threshold point whose
estimated location is $(d_{\rm x}\simeq 4.4, N_{\rm x}\simeq 14)$.
For smaller $d$ and larger $N$, the cuspless fixed point that leads to
dimensional reduction is destabilized by a cuspy fixed point for some
$N=N_\cusp(d)$. This is a rather usual phenomenon, where two fixed
points exchange their stability by crossing. For larger $d$ and
smaller $N$, the critical cuspless fixed point annihilates with an
unstable cuspless fixed point for some $N=N_\DR(d)$. A new, cuspy,
fixed point then emerges from these merged fixed points through a
boundary-layer mechanism. This unusual phenomenon has some specific
signatures in derivatives of the cumulants of the renormalized random
field.  These signatures appear too small to be detected in the
standard short-range RFIM but can be numerically seen in the RFIM in
the presence of long-ranged interactions and disorder correlations.

\section{Acknowledgement}
I. Balog thanks Campus France for financial support.


\begin{thebibliography}{15}
\expandafter\ifx\csname natexlab\endcsname\relax\def\natexlab#1{#1}\fi
\expandafter\ifx\csname bibnamefont\endcsname\relax
  \def\bibnamefont#1{#1}\fi
\expandafter\ifx\csname bibfnamefont\endcsname\relax
  \def\bibfnamefont#1{#1}\fi
\expandafter\ifx\csname citenamefont\endcsname\relax
  \def\citenamefont#1{#1}\fi
\expandafter\ifx\csname url\endcsname\relax
  \def\url#1{\texttt{#1}}\fi
\expandafter\ifx\csname urlprefix\endcsname\relax\def\urlprefix{URL }\fi
\providecommand{\bibinfo}[2]{#2}
\providecommand{\eprint}[2][]{\url{#2}}

\bibitem{tarjus04}
G. Tarjus and M. Tissier, Phys. Rev. Lett. {\bf 93}, 267008 (2004); Phys. Rev. B {\bf 78}, 024203 (2008).

\bibitem{tissier06}
M. Tissier and G. Tarjus, Phys. Rev. Lett. {\bf 96}, 087202 (2006); ; Phys. Rev. B {\bf 78}, 024204 (2008).

\bibitem{tissier06b}
M. Tissier and G. Tarjus, Phys. Rev. B {\bf 74}, 214419 (2006).

\bibitem{tissier11}
M. Tissier and G. Tarjus, Phys. Rev. Lett. {\bf 107}, 041601 (2011); 
Phys. Rev. B {\bf 85}, 104202 (2012); \textit{ibid} {\bf 85}, 104203 (2012).

\bibitem{baczyk13}
M. Baczyk, M. Tissier, G. Tarjus, and Y. Sakamoto, arXiv:1303.2053 (2013).

\bibitem{aharony76}
A. Aharony, Y. Imry, and S. K. Ma, Phys. Rev. Lett. \textbf{37}, 1364 (1976).

\bibitem{grinstein76}
G. Grinstein, Phys. Rev. Lett. \textbf{37}, 944 (1976).

\bibitem{young77}
A. P. Young, J. Phys. C \textbf{10}, L257 (1977).

\bibitem{parisi79}
G. Parisi and N. Sourlas, Phys. Rev. Lett. {\bf 43}, 744 (1979).

\bibitem{fisher86b}
D. S. Fisher, Phys. Rev. Lett. {\bf 56}, 1964 (1986).

\bibitem{narayan92}
O. Narayan and D. S. Fisher, Phys. Rev. B {\bf 46}, 11520 (1992); Phys. Rev. B {\bf 46}, 11520 (1993).

\bibitem{FRGledoussal-chauve}
P. Le Doussal, K. J. Wiese, and P. Chauve, Phys. Rev. B \textbf{66}, 174201 (2002); Phys. Rev. E \textbf{69}, 026112 (2004).

\bibitem{FRGledoussal-wiese}
P. Le Doussal and K. J. Wiese, Phys. Rev. E \textbf{79}, 051106 (2009). 

\bibitem{tarjus13}
G. Tarjus, M. Baczyk, and M. Tissier, arXiv:1209.3161, to appear in Phys. Rev. Lett. (2013).

\bibitem{fisher85}
D. S. Fisher, Phys. Rev. B {\bf 31}, 7233 (1985).

\bibitem{feldman01}
D. E. Feldman, Int. J. Mod. Phys. B \textbf{15}, 2945 (2001).

\bibitem{ledoussal06}
P. Ledoussal and K.J. Wiese, Phys. Rev. Lett.  {\bf 96}, 197202 (2006).

\bibitem{sakamoto06}
Y. Sakamoto, H. Mukaida, and C. Itoi, Phys. Rev. B \textbf{74}, 064402 (2006).

\bibitem{grinstein76b}
G. Grinstein and  A. Luther, Phys. Rev. B, \textbf{13}, 1329 (1976).

\bibitem{footnote} On the other hand, when studying the critical
  behavior at finite temperature, the initial condition is cuspless as
  avalanches are rounded at all nonzero temperatures. In this case one
  should also include the flow of the renormalized temperature which
  is associated with the presence of a thermal boundary layer as one
  approaches the zero-temperature fixed point. Whether or not the
  system then flows to the cuspless fixed point when $d_\cusp>d>d
  _{\DR}(N)$ is unclear to us (but of limited physical consequence
  anyhow).

\bibitem{amit_potts} 
D.J. Amit, J. Phys. A \textbf{9}, 1441 (1976).

\bibitem{priest_potts}
 R.G. Priest and T.C. Lubensky, Phys. Rev. B  \textbf{13}, 4159 (1976).

\bibitem{lubensky}
T.C. Lubensky, B.I. Halperin, and S.K. Ma, Phys. Rev. Lett. \textbf{32},
292 (1974).

\bibitem{dasgupta} 
C. Dasgupta and B.I. Halperin, Phys. Rev. Lett. \textbf{47}, 1556 (1981).

\bibitem{teitel} 
 S. Teitel and C. Jayaprakash, Phys. Rev. B \textbf{27}, 598 (1983).

\bibitem{jones} 
 D.R.T. Jones, A. Love, and M.A. Moore, J. Phys. C \textbf{9}, 743 (1976).

\bibitem{bailin} 
 D. Bailin, A. Love, and M.A. Moore, J. Phys. C \textbf{10}, 1159 (1977).

\bibitem{halperin} 
B.I. Halperin and T.C. Lubenski, Solid State Commun. \textbf{14}, 997
(1974).

\bibitem{lawrie} 
 I.D. Lawrie and C. Athorne, J. Phys. A \textbf{16}, L587 (1983).

\bibitem{march-russel} 
 J. March-Russel, Phys. Lett. B \textbf{296}, 364 (1992).

\bibitem{diep} 
Magnetic Systems with Competing Interactions, edited by H. T.
Diep ͑World Scientific, Singapore, 1994͒






\end{thebibliography}
\end{document}